\documentclass[reprint,twocolumn,superscriptaddress]{revtex4-1}
\usepackage[utf8]{inputenc}
\usepackage{xcolor}
\usepackage{graphicx, graphics,epsfig}
\usepackage{amssymb,amsfonts,amsmath}
\usepackage{ifthen}

\definecolor{ocre}{RGB}{10,100,185}
\usepackage[colorlinks = true,
            linkcolor = ocre,
            urlcolor  = ocre,
            citecolor = ocre,
            anchorcolor = ocre]{hyperref}

\usepackage{color}   
\usepackage[normalem]{ulem}

\newcommand{\EQ}{\begin{equation}}
\newcommand{\EE}{\end{equation}}
\newcommand{\EQA}{\begin{eqnarray}}
\newcommand{\EEA}{\end{eqnarray}}

\begin{document}
\title{Learning the shape of  protein micro-environments  with  \\ a holographic convolutional neural network}
\author{Michael N. Pun}
\affiliation{Department of Physics, University of Washington, 3910 15th Ave Northeast, Seattle, WA 98195, USA}
\affiliation{Max Planck Institute for Dynamics and Self-organization, Am Fa\ss berg 17, 37077 G\"ottingen, Germany}
\author{Andrew Ivanov}
\affiliation{Department of Physics, University of Washington, 3910 15th Ave Northeast, Seattle, WA 98195, USA}
\author{Quinn Bellamy}
\affiliation{Department of Physics, University of Washington, 3910 15th Ave Northeast, Seattle, WA 98195, USA}
\author{Zachary Montague}
\affiliation{Department of Physics, University of Washington, 3910 15th Ave Northeast, Seattle, WA 98195, USA}
\affiliation{Max Planck Institute for Dynamics and Self-organization, Am Fa\ss berg 17, 37077 G\"ottingen, Germany}
\author{Colin LaMont}
\affiliation{Max Planck Institute for Dynamics and Self-organization, Am Fa\ss berg 17, 37077 G\"ottingen, Germany}
\author{Philip Bradley}
\affiliation{Fred Hutchinson Cancer Research Center, 1100 Fairview Ave N, Seattle, WA 98109, USA}
\author{Jakub Otwinowski}
\affiliation{Dyno Therapeutics, 343 Arsenal St, Suite 101, Watertown, MA 02472}
\author{Armita Nourmohammad}
\email{Correspondence should be addressed to: armita@uw.edu}
\affiliation{Department of Physics, University of Washington, 3910 15th Ave Northeast, Seattle, WA 98195, USA}
\affiliation{Max Planck Institute for Dynamics and Self-organization, Am Fa\ss berg 17, 37077 G\"ottingen, Germany}
\affiliation{Fred Hutchinson Cancer Research Center, 1100 Fairview Ave N, Seattle, WA 98109, USA}
\affiliation{Department of Applied Mathematics, University of Washington, 4182 W Stevens Way NE, Seattle, WA 98105, USA}

\begin{abstract}
Proteins play a central role in biology from immune recognition to brain activity. While major advances in machine learning have improved our ability to predict protein structure from sequence, determining protein function from structure remains a major challenge. Here, we introduce Holographic Convolutional Neural Network (H-CNN) for proteins, which is a physically motivated machine learning approach to model amino acid preferences in protein structures. H-CNN reflects physical interactions in a protein structure and recapitulates the functional information stored in evolutionary data. H-CNN accurately predicts the impact of mutations on protein function, including stability and binding of protein complexes. Our interpretable computational model for protein structure-function maps could guide design of novel proteins with desired function.
\end{abstract}

\maketitle

Proteins are the machinery of life. They facilitate the key processes that drive living organisms and only rely on twenty types of amino acids to do so. The physical arrangement of a protein's amino acids dictates how it folds and interacts with its environment. While this compositional nature gives rise to the diversity of existing proteins, it also makes determining protein function from structure a complex problem.

With the growing amount of data and computational advances, machine learning has come to the forefront of protein science, especially in predicting structure from sequence~\cite{AlQuraishi2019-rs,Gao2020-zi,Jumper2021-hl,Baek2021-gq,Bouatta2021-ez}. However, the problem of how protein function is determined from its sequence or structure still remains a major challenge.

Techniques from natural language processing are used to determine functional motifs in protein sequences by allowing residues far away in sequence to form information units about function~\cite{Alley2019-pp,Rao2019-dv,Madani2020-xb,Bepler2021-uo,Rives2021-sr,Hie2021-aa,lin2022language}. However, since protein function is closely related to protein structure, models trained to map protein sequences to function must account for the complex sequence-structure map.

Despite AlphaFold's remarkable success at predicting protein folding, it still struggles to determine the effect of  mutations on the stability and function of a protein~\cite{pak_using_2021}. Nonetheless, it is suggested that AlphaFold has learned an effective physical potential to fold proteins, and therefore,  it could be used to characterize the effect of mutations or general protein function~\cite{roney_state---art_2022}. Given the availability of high-resolution tertiary structures, obtained either experimentally or computationally, the information on atomic coordinates in a 3D structure of a protein can be used to learn a direct map to function. Indeed, models aware of the geometry of protein 3D structure that  attempt to solve the inverse folding problem, i.e., designing a sequence that folds into a desired structure, can be used to reliably infer the functional effect of mutations in a protein sequence~\cite{hsu_learning_2022}, or even engineer diverse sequences that have a desired function~\cite{Dauparas2022-fu}.

Recent work in the field of molecular dynamics (MD) has shown the power of geometry-aware machine learning at inferring precise inter-atomic force fields~\cite{batzner_e3-equivariant_2022,musaelian_learning_2022,satorras_en_2021,tholke_equivariant_2022,schutt_equivariant_2021,haghighatlari_newtonnet_2022,gasteiger_gemnet_2021,batatia_design_2022}. Compared to the geometry-aware protein structure models~\cite{Jumper2021-hl,Baek2021-gq,hsu_learning_2022,Dauparas2022-fu}, the MD models  use more complex geometric features, resulting in more expressive, yet physically interpretable models of molecular interactions. 

\begin{figure*}[t!]
\includegraphics[width=\textwidth]{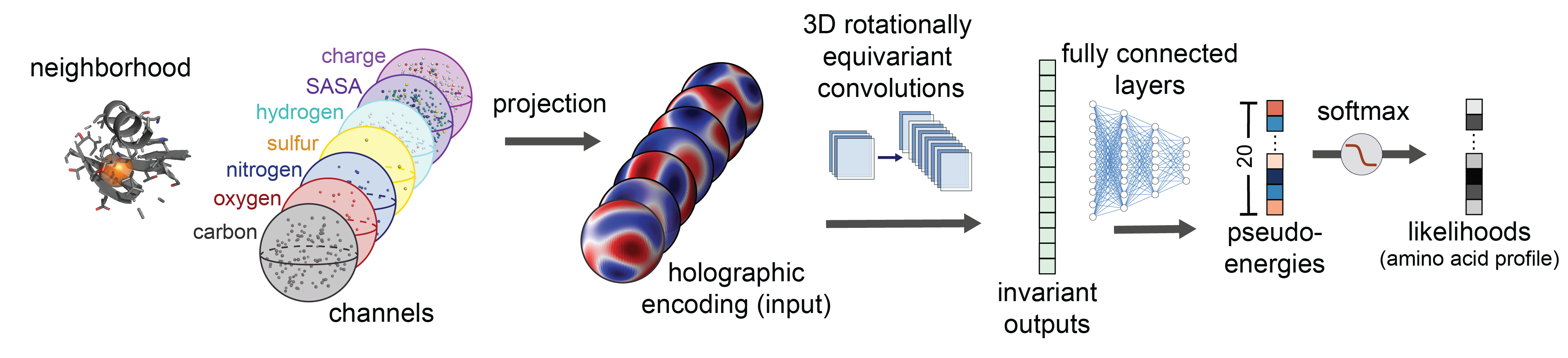}
 \caption{
{\bf Schematic of Holographic Convolutional Neural network (H-CNN) for protein micro-environments.} A neighborhood within a  radius of 10\AA\, around a focal amino acid  (masked in orange) in a protein structure is separated into its constituent atomic and chemical channels. The information in these channels is encoded in a rotationally equivariant form, using 3D Zernike polynomials, which defines holograms in spherical Fourier space. These holographic encodings are processed by a  rotationally equivariant convolutional neural network (Clebsch-Gordan net~\cite{Kondor2018-tr}). The invariant features of the network layers are then collected and processed through fully connected feed-forward layers to determine the preferences, i.e., statistical weights (pseudo-energies) and probabilities, for different amino acids residing  at the center of the input neighborhood. The set of predicted probability vectors across all 20 amino acids defines an amino acid profile. The network is trained by learning the categorical classification task with a softmax cross-entropy loss on a one-hot label of the neighborhood determined by the true central residue in the protein structure. A more detailed network architecture is presented in Fig.~S2.
\label{Fig:1}}
\end{figure*}

Here, we introduce  holographic convolutional neural network (H-CNN) to learn physically grounded structure-to-function maps for proteins. H-CNN learns local representations of protein structures by encoding the atoms  within protein micro-environments in a  spherical Fourier space as  holograms, and processes  these  holographic projections via a 3D rotationally equivariant convolutional neural network (Fig.~\ref{Fig:1})~\cite{Cohen2016-jk,Kondor2018-tr,Thomas2018-uz}. The resulting model respects rotational symmetry of protein structures and characterizes effective inter-amino-acid potentials in protein micro-environments.

We train H-CNN on protein structures available in the Protein Data Bank (PDB)~\cite{Chaudhury2010-mo}, and perform the supervised task of predicting the identity of an amino acid from its surrounding atomic neighborhood with a high accuracy and computational efficiency. The amino acids  that H-CNN infers to be  interchangeable have similar  physico-chemical properties, and the pattern is consistent with substitution patterns in evolutionary data. The H-CNN model encodes a more complete set of geometric features of protein structures compared to the other geometry-aware  models of proteins~\cite{hsu_learning_2022,lin2022language,Dauparas2022-fu}. Therefore, it can predict  protein function, including binding and stability effects of mutations,  only based on local atomic compositions in a protein structure. Our results showcase that principled geometry-aware machine learning  can lead to  powerful and robust models that provide insight into the biophysics of protein interactions and function, with a potential for protein design.

\begin{figure*}[t!]
\includegraphics[width=0.9\textwidth]{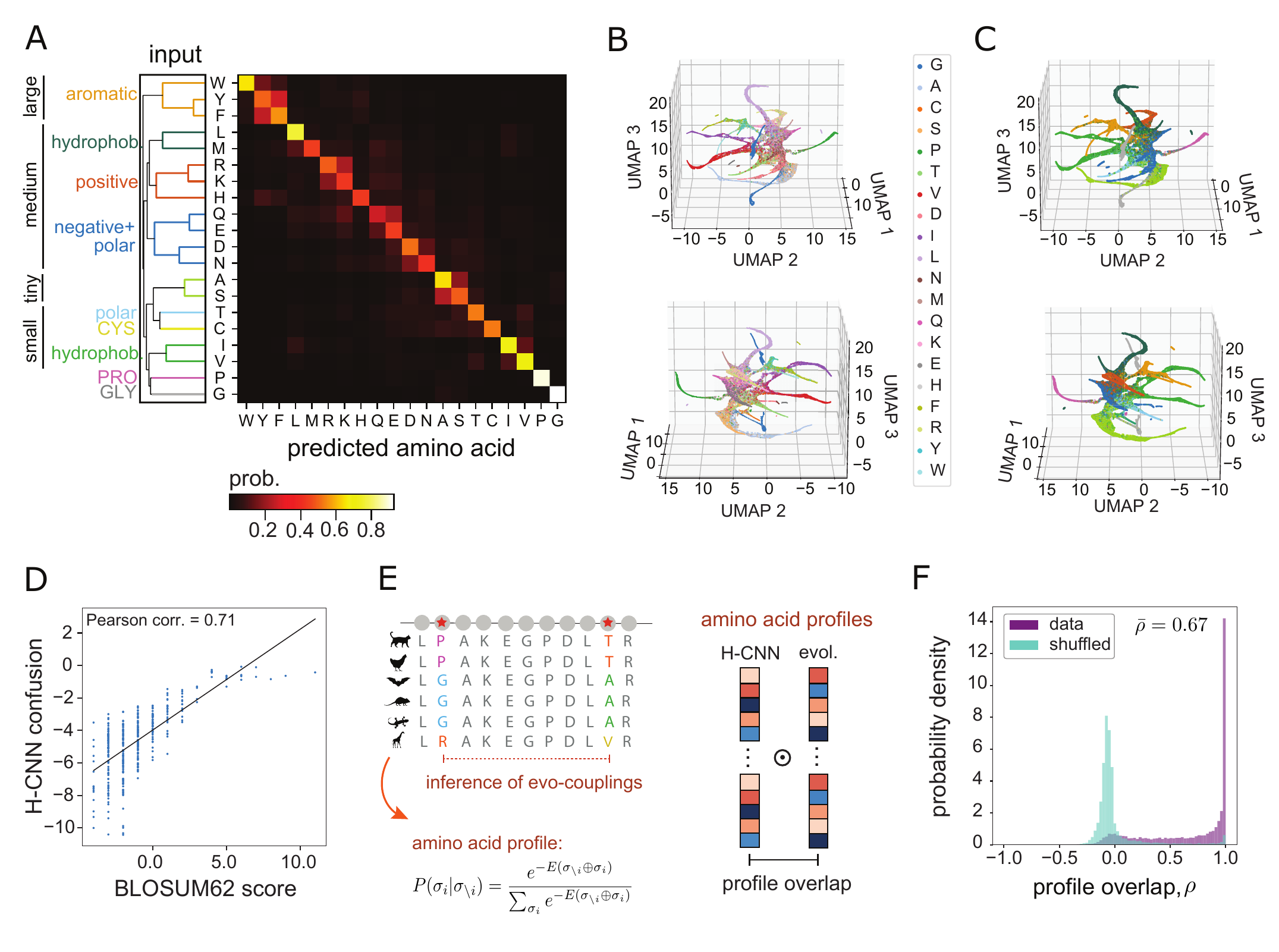}
\caption{
{\bf H-CNN predicts amino acid preferences in protein micro-environments.} {\bf (A)} The confusion matrix for  amino acid predictions with  H-CNN shows the mean H-CNN predicted probabilities of each of the twenty amino acids (output) conditioned on a specific central amino acid (input). Overall prediction accuracy is 68\%.
The hierarchical clustering for these predictions reflects known similarities in size and physico-chemical properties of amino acids. {\bf (B,C)} Low dimensional projections of the prediction outputs (3D UMAPs) are shown. UMAPs are annotated by  {\bf (B)}  the amino acid types, and {\bf (C)} the physico-chemical clusters in (A), with panels showing a different view of the  UMAP in each case. Neighborhoods are closely clustered by amino acid types (B), and are spatially arranged based on the physico-chemical properties of the side-chains (C); colors  in (C) are consistent with (A). 
{\bf(D)} Amino acid confusion in (A) correlates with the substitutability of amino acids in natural proteins as determined by the BLOSUM62 matrix; $71\%$ Pearson correlation. {\bf (E)} Schematic shows how evolutionary covariation of amino acids in multiple sequence alignments of  protein families  can be used to fit Potts models (EV-couplings~\cite{Hopf2018-li}) to characterize the probability of a given amino acid, given the rest of the sequence (left); see Methods for details. To compare evolutionary and H-CNN predictions for site-specific amino acid profiles,  the profile overlap is computed as the centered cosine similarity between the predicted probability profiles (right); see Methods. 
{\bf (F)} The profile overlaps  are strongly peaked around one,  implying perfect overlap in data (purple); the average profile overlap across 11,221 sites from a total of 67 protein families is $\bar\rho=0.67$.  The H-CNN predictions are notably different for the shuffled data, for which the profile overlap peaks near zero (cyan), with an average of 0.002. }
\label{Fig:2}
\end{figure*}

\section{Results}

\subsection*{Model}
We define the micro-environment surrounding an amino acid as the identity and the 3D coordinates of atoms within a radius of 10~\AA\,  of the focal amino acid's $\alpha$-carbon; this neighborhood excludes atoms  from the focal amino acid.  

A common approach to encode such atomic neighborhoods for computational analysis is to voxelize the coordinates, which is a form of binning in 3D~\cite{Torng2017-aq,Shroff2020-bu}. However, this approach distorts the information, since the voxel boundaries are arbitrary---too large voxels average over many atoms, and too small voxels lead to very sparse data. 
 
The other obstacle for modeling such data is more fundamental and related to the rotational symmetries in encoding a protein structure neighborhood.  A given neighborhood can occur in different orientations within or across proteins, and a machine learning algorithm should account for such rotational symmetry.  One approach known as data augmentation,   mainly used in image processing, trains an algorithm on many examples of an image in different orientations and locations. Data augmentation is computationally costly in 3D, and it is likely to result in a model of amino acid interactions that depends on the neighborhood’s orientation, which is a non-physical outcome. Another approach is to orient the amino acid neighborhoods based on a prior choice (e.g., along the backbone of the protein)~\cite{Torng2017-aq,Shroff2020-bu}. However, this choice is somewhat arbitrary, and the specified orientation of the protein backbone could inform the model about the identity of the  focal amino acid. 
 
To overcome these obstacles, we introduce holographic convolutional neural networks (H-CNNs) for protein micro-environments. First, we encode the point clouds of different atoms in an amino acid neighborhood using  3D Zernike polynomials as spherical basis functions (Fig.~\ref{Fig:1} and Methods).  3D Zernike polynomials can be used to expand any function in three dimensions along angular and  radial bases and  can uniquely represent the properties of the encoded object in a spherical Fourier space, given enough terms in the Fourier series. Conveniently, the angular component of the Zernike polynomials are spherical harmonics, which form an equivariant basis under rotation in 3D. Rotational equivariance is the property that if the input  (i.e., atomic coordinates of an amino acid's neighborhood) is rotated, then the  output is transformed according to a linear operation determined by the rotation (Fig.~S1). As a result, these Zernike projections enable us  to encode the atomic point clouds from a protein structure without having to align the neighborhoods. Zernike projections in  spherical Fourier space can be understood as a superposition of spherical holograms of an input point cloud, and thus, we term this operation as {\em holographic encoding} of protein  micro-environments; see Fig.~\ref{Fig:1} and Fig.~S2 and Methods for details. 

The  holograms  encoding protein neighborhoods are input to a type of convolutional neural network (CNN). This network is trained on the supervised task of predicting the identity of a focal amino acid from the surrounding atoms in the protein's tertiary structure. Conventional CNNs average over spatial translations and can learn features in the data that may be in different locations (i.e., they respect translational symmetry).  For the analysis of protein neighborhoods we need to infer models that are insensitive to the orientation of the data (i.e., that respect 3D rotational symmetry of the point clouds in a protein neighborhood). Recent work in machine learning has expanded CNNs to  respect physical symmetries beyond translations~\cite{Cohen2016-jk,Kondor2018-tr,Thomas2018-uz}. For 3D rotations, generalized convolutions use spherical harmonics, which arise from the irreducible representations of the 3D rotation group SO(3)~\cite{tung_group_1985}. For our analysis, we use Clebsch-Gordon neural networks~\cite{Kondor2018-tr}, in which the linear and the nonlinear operations of the network layers have the property of rotational equivariance (Methods); see Fig.~S2 for detailed information on network architecture and Fig.~S3 and Table~S1 for details on hyper-parameter tuning and training of the network.  

Taken together, the H-CNN shown in Figs.~\ref{Fig:1},~S2, takes as input holograms that encode the spatial composition of different atoms (Carbon, Nitrogen, Oxygen, Sulfur, Hydrogen) and physical properties such as charge and solvent accessible surface area (SASA). The input is processed by a 3D rotationally equivariant CNN to learn statistical representations for protein neighborhoods. We train this H-CNN as a classifier on  protein neighborhoods, collected from tertiary structures from the PDB, and use the trained network to  quantify the preferences for different amino acids in a given structural neighborhood; see Methods for details on data pre-processing.

\subsection*{H-CNN reveals physico-chemical   properties  of amino acids, consistent with evolutionary variation}
H-CNN predicts the identity of an amino acid from its surrounding micro-environment with 68\% accuracy (Fig.~\ref{Fig:2}). The  accuracy of H-CNN is comparable to state-of-the-art approaches with conventional CNNs that voxelize and orient the data along the backbone of a central amino acid, while using a smaller atomic micro-environement for performing this classification task~\cite{Torng2017-aq,Shroff2020-bu}; see Table~S2 for a detailed comparison of models. Notably, restricting the training of H-CNN to the subspace of models that are rotationally equivariant leads to a substantial speedup in the training of H-CNN compared to the conventional techniques~\cite{Torng2017-aq,Shroff2020-bu}. Moreover, H-CNN is more accurate than other symmetry-aware approaches for molecular modeling, while using an order of magnitude fewer parameters~\cite{boomsma_spherical_2017, weiler_3d_2018}; see Methods and Table~S2 for a detailed comparison of models. 

H-CNN predicts the conformationally unique amino acids of Glycine and Proline  with over $90\%$ accuracy. Meanwhile, amino acids with typical side-chains cluster based on  their sizes and the physico-chemical properties of the side-chains including aromatic, hydrophobic, and charged groupings (Fig.~\ref{Fig:2}A). The inferred amino acid preferences cluster well according to the input amino acid type (true label) in the low-dimensional UMAP representation~\cite{mcinnes_umap_2018-1}, and  amino acids with similar physico-chemical  properties cluster in nearby regions in the UMAP (Fig.~\ref{Fig:2}B,C).

 H-CNN predictions reflect amino acid preferences seen in evolutionary data, even though the network is not trained on multiple sequence alignments (MSAs) of protein homologs. Specifically, the interchangeability of amino acids that H-CNN predicts is 71\% correlated with  the substitution patterns in evolutionary data, represented by the BLOSUM62 matrix  (Fig.~\ref{Fig:2}D). In addition, the amino acid preferences predicted by H-CNN at each site are consistent with evolutionary preferences inferred from the covariation of residues in multiple sequence alignments of protein families~\cite{Morcos2011-ku,Hopf2017-ul, Hopf2018-li}; see Fig~\ref{Fig:2}E,F and Methods for details. 

 Ablation studies further reveal that the H-CNN's  processing of information corresponds to physical intuition. Removing information about SASA  or charge from the input data results in roughly a 10\% drop in classification accuracy; see Fig.~S4 and the discussion on ablation studies in the Methods.  Information from SASA  mostly impacts the network's ability to predict hydrophobic amino acids, with some hydrophilic amino acids (R,K,E) also impacted. When charge is removed, the network demonstrates worse predictions on charged and polar amino acids most notably R, C, N, and E.

\begin{figure*}[t!]
\includegraphics[width=\textwidth]{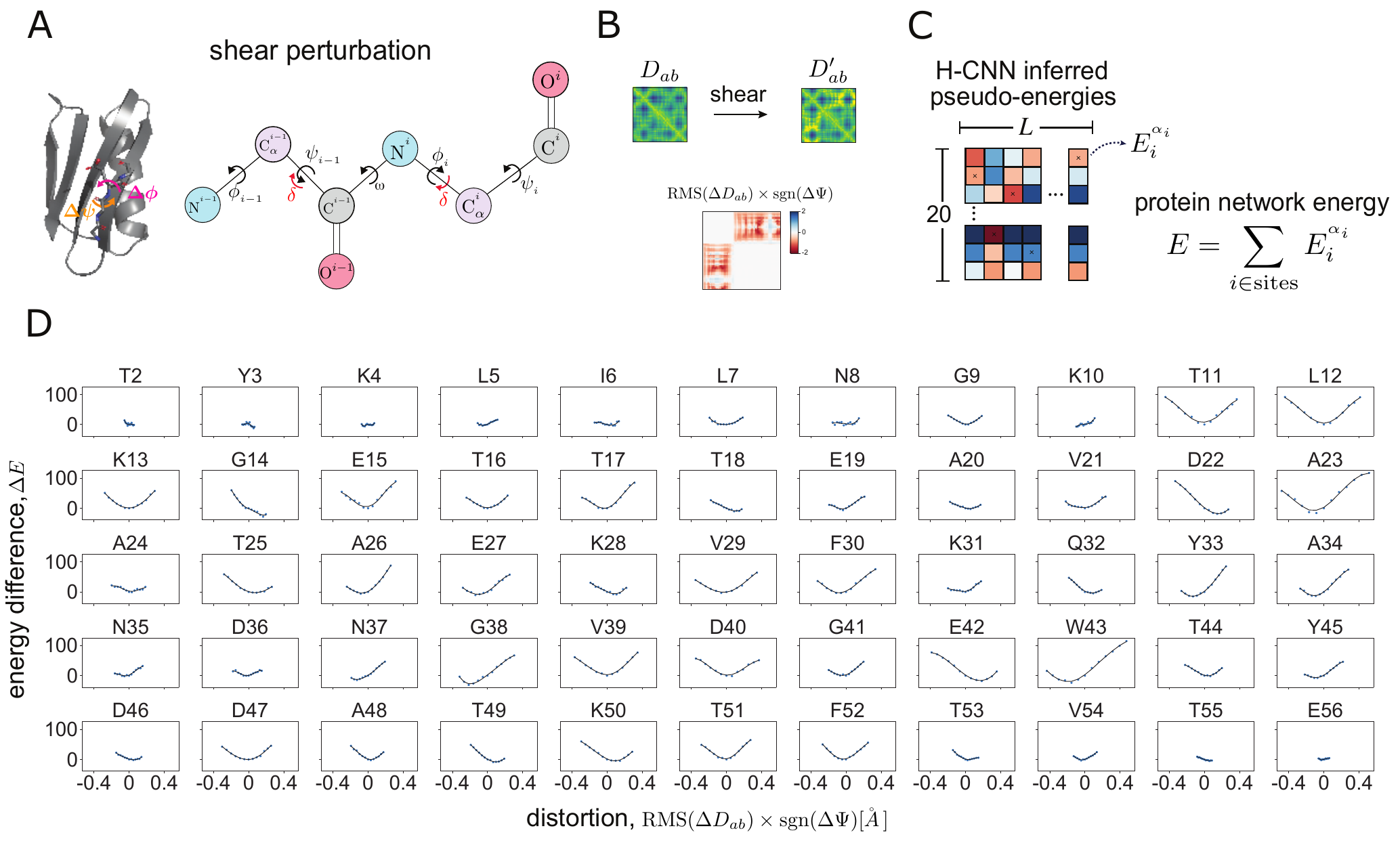}
\caption{{\bf Response of H-CNN predictions to physical distortions in protein structures.} {\bf (A)} The schematic shows shear perturbation in a protein backbone by an angle $\delta$ at site $i$ as a rotation of side-chains around the backbone by the angles $[\phi_i,\psi_{i-1}] \rightarrow[ \phi_i + \delta, \psi_{i-1} - \delta ]$~\cite{Chaudhury2010-mo}. {\bf (B)} Shearing changes the pairwise distance matrix $D_{ab}$ between all atoms in a protein structure. The total physical distortion is computed as the root-mean-square of changes in the pairwise distances that are less than 10 \AA\, (i.e., residues within the same neighborhood), multiplied by the sign of the change in the angle $\psi$. {\bf (C)} For a given perturbation, the network energy $E$ is determined by the sum of pseudo-energies of the wild-type amino acid at all sites in the protein, and the change in this quantity by shearing $\Delta E$ measures the tolerance of a structure to a given perturbation.  {\bf (D)} Panels show the change in the network energy in response to the structural distortion by shear perturbation at all sites in protein G, with the amino acid type and the site number indicated above each panel.
 \label{Fig:3}}
\end{figure*}

\subsection*{H-CNN learns an effective physical potential for protein micro-environments}
Since H-CNN is trained to predict the most natural amino acid given its neighborhood, it should also be able to recognize an unnatural protein configuration. To test this hypothesis, we characterize the response of the H-CNN  predictions to physical distortions in native atomic micro-environments. We introduce  distortions through local {\em in silico} shear perturbation of the protein backbone at a given site $i$ by an angle $\delta$, resulting in a transformation of the backbone angles by $\phi_i \rightarrow \phi_i + \delta, \,  \psi_{i-1} \rightarrow \psi_{i-1} - \delta$ (Fig.~\ref{Fig:3}A and Methods); this perturbation is local with minimum downstream effects~\cite{Chaudhury2010-mo}. We measure the distortion of the protein structure due to shear by calculating the change in the root-mean-square deviation in the pairwise distances of all atoms of the perturbed protein structure relative to that of the wild-type ($\text{RMS}\Delta D_{ab},\,  \text{for all pairs of atoms}\, (a,b)$); Fig.~\ref{Fig:3}B.  

 We measure the response of the protein to shear perturbation by analyzing the change in the logits produced by  H-CNN, which we term pseudo-energies $E^{\alpha}_i$ (Fig.~\ref{Fig:1}). Specifically, for a given distorted structure, we  re-evaluate the  pseudo-energy of each amino acid in the protein, and define the total H-CNN predicted energy by summing over the pseudo-energies of all  the amino acids in a protein (Fig.~\ref{Fig:3}C). The change  in the predicted energy of a protein due to distortion (relative to the wild-type) $\Delta E$ is a measure of H-CNN's response to  a given  perturbation. A positive $\Delta E$ indicates an unfavorable change in the protein structure.   

In protein G, the change in the predicted energy $\Delta E$ as a function of distortion in the structure $\text{RMS}\Delta D_{ab}$ due to shearing at different sites reveals two trends (Fig.~\ref{Fig:3}D). First, the protein network energy appears to respond locally quadratically to  perturbations. Second, perturbations generally result in higher protein network energy, corresponding to a less favorable protein micro-environment. Taken together, by training on a classification task and by constraining the network to respect the relevant rotational symmetry,  H-CNN has learned an effective physical  potential for protein micro-environments in which the native structure is generally more favorable  and robust to local perturbations (i.e., it is at the energy minimum). 

This observation of a minimum energy extends beyond the wild-type sequence when biophysically similar amino acids are substituted in the energy sum (Methods and Fig.~S5A). Notably this pattern appears not to be just an artifact of the structure since the minimum disappears when random amino acids are used to calculate the network energy (Fig.~S5B).

\begin{figure*}[t!]
\includegraphics[width=\textwidth]{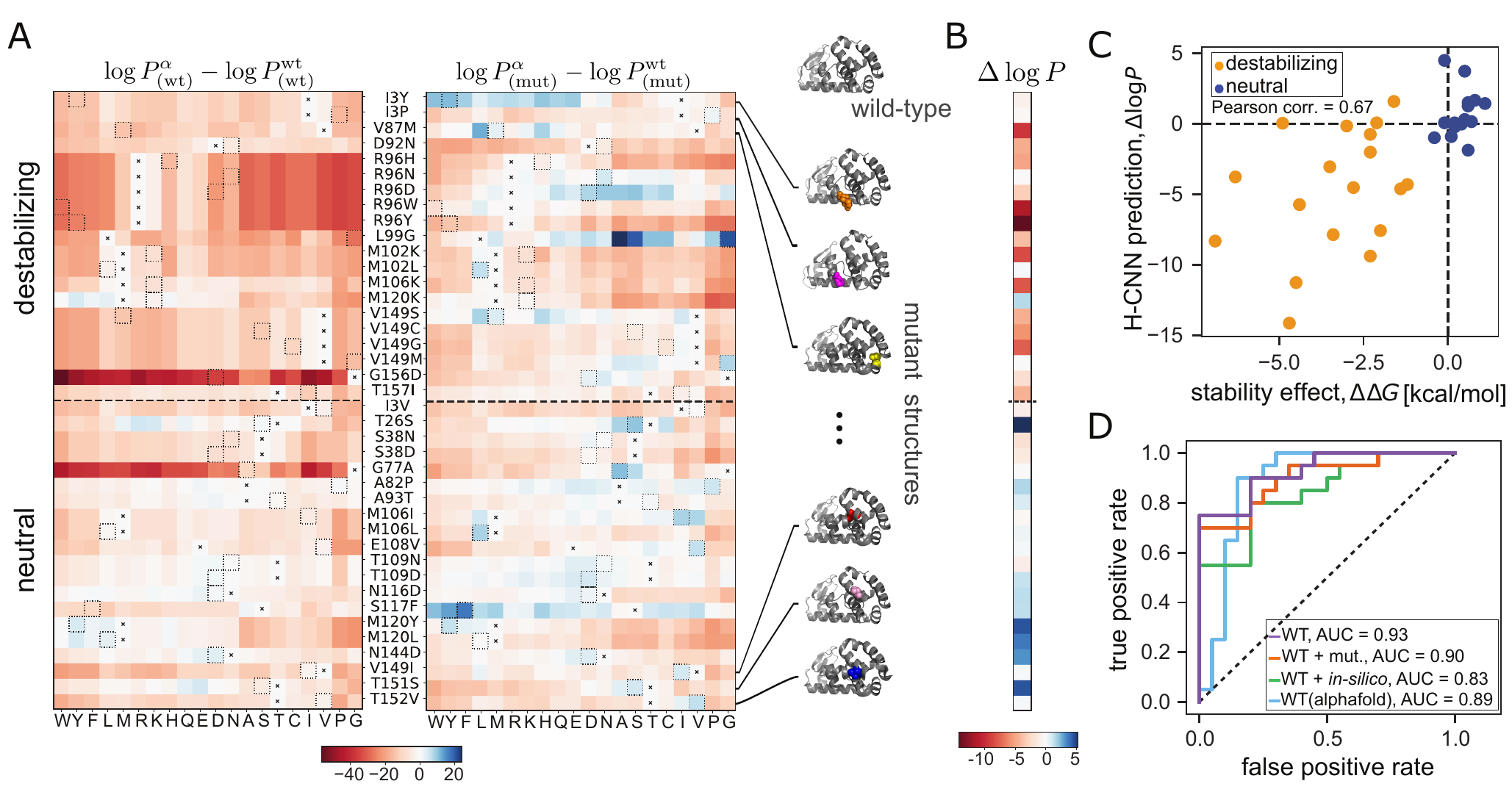}
\caption{{\bf Predicting the stability effect of mutations in T4 lysozyme with H-CNN.}
{\bf (A)}    Heatmaps of H-CNN predicted log probability of different amino acids (columns) relative to  that of the wild-type amino acid for 40 variants with single amino acid substitution from the wild-type (rows). For each variant (row), the position and the identity of the wild-type amino acid and the mutation are denoted between the two heatmaps as: wild-type, site number, mutation. The left panel shows the predictions using the wild-type protein structure (subscript (wt)), while the right panel shows the predictions using the structure of the specified mutant at each row (subscript (mut)). In each row the wild-type amino acid is indicated by an $\times$, and a dotted box shows the amino acid of the mutant.  {\bf (B)}  The H-CNN predicted log-probability ratio  $\Delta \log P = \log P^{\rm mut}_{\rm (mut)} /   P^{\rm wt}_{\rm (wt)}$ of the mutant amino acid on the mutant structure with respect to the wild-type amino acid on the wild-type structure is shown for all variants. The predicted  ratios for destabilizing mutations are negative, while those for the neutral / beneficial mutations are positive. {\bf(C)} The H-CNN predicted log-probability ratio $\Delta \log P$ shown against the experimentally evaluated $\Delta\Delta G$ for  the  stability effect of mutations in each protein structure; Pearson correlation of  67\%.
{\bf (D)} The H-CNN predictions for the relative log-probabilities $\Delta \log P_\text{wt}$ using  the wild-type structure only  are shown against the experimentally measured $\Delta \Delta G$ values for 310 single point mutation variants of  T4 lysozyme.   Mean $\Delta\Delta G$ was used when multiple experiments reported values for the same variant. The colors  show the density of points as calculated via Gaussian kernel density estimation. The predictions are accurate with correlations indicated in the panel.
\label{Fig:4}
}
\end{figure*}

\begin{figure*}[t!]
\includegraphics[width=0.8\textwidth]{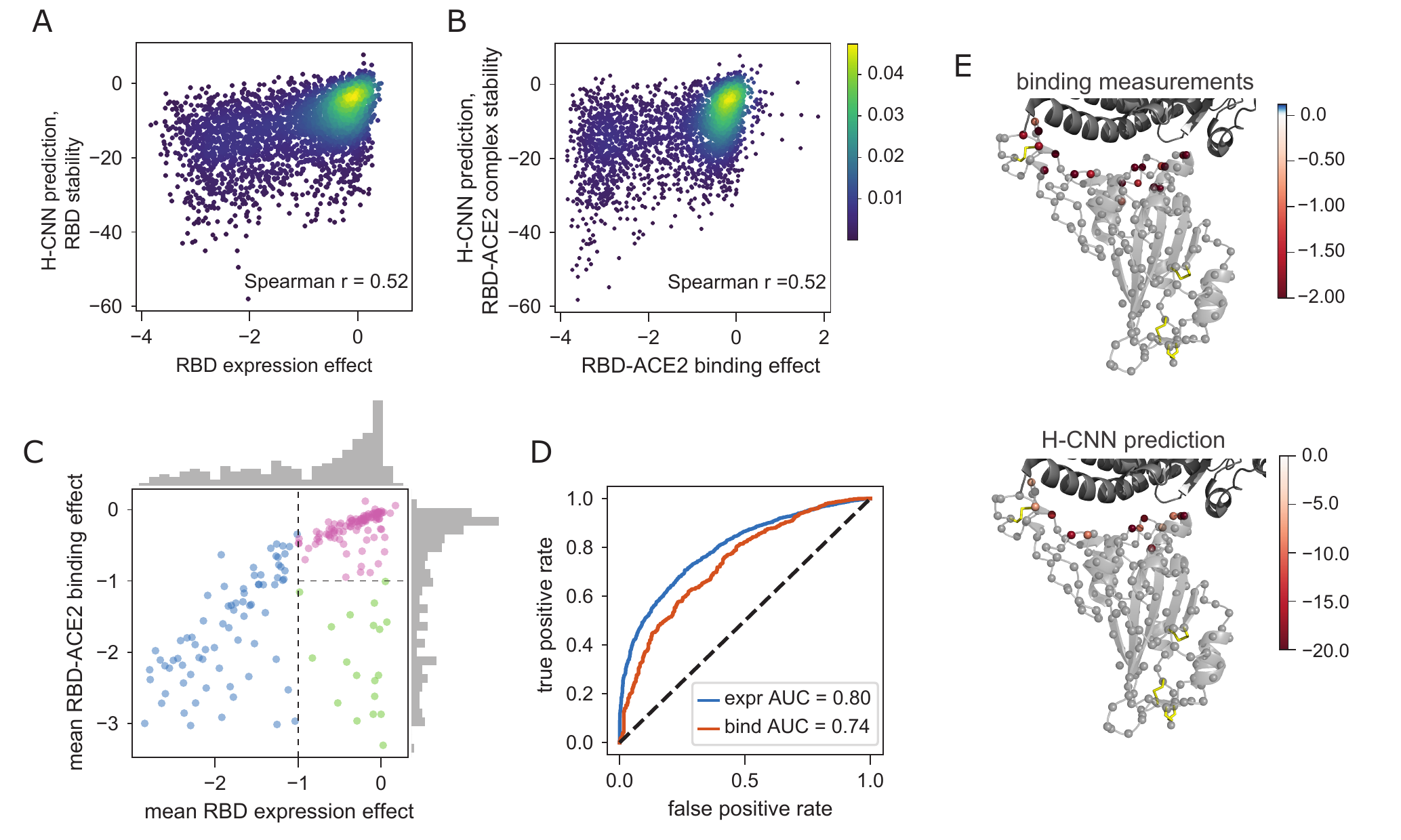}

\caption{{\bf Predicting the stability and binding of the RBD protein of SARS-CoV-2 with H-CNN.}
{\bf(A)} The density plot shows  H-CNN predictions for the RBD stability, using the isolated protein structure of RBD, against  the mutational effects on the RBD expression from the DMS experiments; Spearman correlation $r = 0.52$. {\bf (B)} The density plot shows  H-CNN predictions  for the RBD binding to the ACE2 receptor, using the co-crystallized  RBD-ACE2 protein structure, against the  DMS measurements for mutational effects on binding; shared color bar for (A) and (B). {\bf (C)} The mean effect of mutations at each site on the RBD-ACE2 binding is shown against the mean effect on the RBD expression. The histograms show the corresponding distribution of effects across sites along each axis. The categories are shown: (i) sites that are intolerant to mutations due to destabilizing effect, i.e., low expression (blue), (ii) sites that are tolerant of mutations for expression but not binding (green), and (iii) sites that are tolerant of mutations for both expression and binding (pink).
{\bf (D)} Blue:  true positive vs. false positive rate (ROC curve) for classification of amino acid mutations into stable (expr $> -1$) vs. unstable (expr $<-1$), based on the H-CNN predictions using the isolated  RBD structure;  AUC =0.8. Red: the ROC curve for mutation classification into bound (bind $>-1$) vs. unbound (bind $<-1$), based on  the H-CNN predictions using  the co-crystallized  RBD-ACE2 structure; AUC= 0.74. {\bf (E)} The effect of mutations on binding from the DMS experimental data for the green sites in (C) (top) and the corresponding H-CNN predictions from the RBD-ACE2  structure complex for sites identified by H-CNN in Fig.~S10 to be tolerant of mutations for stability but not binding   (bottom) are shown throughout the structure. 
}
\label{Fig:5}
\end{figure*}

\subsection*{H-CNN predicts  effect of mutations on protein stability}

Characterizing amino acid preferences in a  protein neighborhood is closely related to the problem of finding the impact of mutations on protein function. Here, we test the accuracy of H-CNN in predicting the stability effect of mutations in 40 different variants of  the T4 lysozyme protein.  Each of these variants is one amino acid away from the wild-type, with variations spanning 23 residues of the protein. Notably,  the tertiary structure of the wild-type T4 lysozyme protein as well as the 40 mutants are available through different studies~\cite{Grutter1987-eo, Gray1987-hk, Hurley1992-dp, Mooers2009-ik, Dixon1992-ye, Weaver1989-fo, Dao-pin1991-gs, Matsumura1988-gz, Lipscomb1998-yq, Anderson1993-ap, Wray1999-ox, Xu2001-vy, Mooers2003-eq, Nicholson1988-uu, Nicholson1991-wr, Gassner1999-uq, Pjura1993-mx, matthews_enhanced_1987, Torng2017-aq}; see Table~S3 for details on these mutants. 

H-CNN  predicts that the wild-type amino acids are the most favorable in the wild-type structure, while the mutant amino acids are generally more favorable in the mutant structures, regardless of their stabilizing effects (Fig.~\ref{Fig:4}A). These variant-specific preferences are not surprising since the folded protein structure can relax to accommodate for  amino acid changes, resulting in a structural neighborhood that is more consistent with the statistics of the micro-environments around the mutated amino acid than that of the wild-type. However, the confidence that H-CNN has in associating an amino acid with a given structural neighborhood can change depending on the stability effect of the mutation. The log-ratio of the H-CNN inferred probability for the mutant amino acid in the mutant structure  versus that of the wild type amino acid in the wild type structure,  $\Delta \log P = \log P_\text{mut} / P_\text{wt}$ can provide an approximation to the $\Delta \Delta G$ associated with the stability of a mutation (Methods). 
 
The inferred H-CNN predicted log-probability ratio is generally negative for destabilizing mutations, and non-negative for  neutral/weakly beneficial mutations  (Fig.~\ref{Fig:4}B). Previously, a structure-based CNN model with voxelized protein structures has shown a similar qualitative result~\cite{Torng2017-aq}. Further quantitative analysis shows that the log-probability ratio is 67\% correlated with the experimentally evaluated $\Delta\Delta G$ values for these variants (Fig~\ref{Fig:4}C). Moreover, the receiver-operating-characteristic (ROC) curves in Fig.~S6A show that the  log-ratio of amino acid probabilities can reliably discriminate between destabilizing and neutral mutations, with an area under the curve (AUC) of 0.90. 

The availability of tertiary structures for a large number of variants is a unique feature of this dataset, and in most cases such structural resolution is not accessible.  To overcome  this limitation and predict the stability effect of mutations by relying on the wild-type structure alone, we used PyRosetta to relax the wild-type T4 lysozyme structure around a specified amino acid change~\cite{Chaudhury2010-mo} (Methods).  We find that the log-probability ratios $\Delta \log \tilde P$ estimated based on these {\em in silico} relaxed mutant structures are mostly negative (non-negative) for destabilizing (neutral) mutations (Fig.~S7) and are correlated with the stability effect of  mutations  $\Delta\Delta G$ (Fig.~S7).  However, structural relaxation can add noise to the data, causing the protein micro-environments to deviate from the  natural structures that H-CNN is trained on. Thus, using the {\em in silico} relaxed structures slightly  reduces the discrimination power of our model between deleterious and near-neutral mutations (AUC = 0.83); see Fig.~S6A. 
  
In contrast, the preferences estimated  based on the wild-type structure only  can  discriminate between destabilizing and neutral mutations very well, even though  most mutations are inferred to be deleterious with respect to the wild-type  (AUC~=~0.93 in Fig.~S6). In other words, by using the wild-type structure only, our model can predict the relative stability effect of mutations correctly but not the sign of $\Delta \Delta G$ (Fig.~S6, S7). Indeed, our inferred log-probability ratios based on the wild-type structure show a substantial  correlation of 64\% (Pearson correlation)  with the stability effect of a much larger set of 310 single point mutants~\cite{stourac_fireprotdb_2021}, for which protein structures are not available (Fig~\ref{Fig:4}D). 

When no experimentally determined structure is available, computationally resolved protein structures from AlphaFold can also be used to predict the stability effect of mutations. The H-CNN predictions using the template-free AlphaFold2 predicted structure of T4 lysozyme wild-type sequence display substantial discrimination ability between destabilizing and near-neutral mutations (Fig.~S6) and are correlated with the mutants' $\Delta\Delta G$ values (Figs.~S6,~S7).

\subsection*{H-CNN predicts fitness effect of mutations for binding of SARS-CoV2  to the ACE2 receptor}
Recent deep mutational scanning (DMS) experiments measured the effect of thousands of mutations in the receptor-binding domain (RBD) of SARS-CoV-2 on the folding of the RBD (through expression measurements) and its binding to the human Angiotensin-Converting Enzyme 2 (ACE2) receptor~\cite{Starr2020-rr,Starr2022-lp}.

H-CNN  can be used to predict the effect of mutations  on RBD, either in isolation or bound to  the ACE2 receptor. 
The former can be interpreted as  the effect of mutations on the stability of RBD, which is measured by the expression of the folded domain in the experiments~\cite{Adams2016-ac,Starr2020-rr,Starr2022-lp}, while the latter can be used to characterize  amino acid preferences for binding at the RBD-ACE2  interface. Fig.~\ref{Fig:5}A,B shows that the H-CNN predictions are correlated with the stability and binding  measurements in the DMS experiments from ref.~\cite{Starr2022-lp}; site specific effects are depicted in Figs.~S8,~S9. 

The average effect of mutations on expression and binding can define three categories of sites and/or mutations (Fig.~\ref{Fig:5}C): (i) sites that are intolerant to mutations (due to destabilizing effects) and show a substantially reduced expression of mutants (blue), (ii)  sites that are tolerant of mutations for expression but not binding (green), and (iii) sites that are tolerant of mutations for both expression and binding (pink). Using the  isolated  structure of  RBD,  H-CNN can well classify mutations according to their stability effect (AUC~=~0.8; Fig.~\ref{Fig:5}D). Similarly, with the structure of the RBD-ACE2 complex, H-CNN can classify mutations according to their tolerance for binding (AUC~=~0.74; Fig.~\ref{Fig:5}D). 

Expectantly, the sites that are tolerant of mutations for expression but not binding (green category from the DMS data in Fig.~\ref{Fig:5}C) are located at the interface of the RBD-ACE2 complex, and H-CNN correctly predicts this composition (Fig.~\ref{Fig:5}E, Fig.~S10). The overall impact of mutations on binding for these sites is shown in Fig.~\ref{Fig:5}E.

Identifying candidate sites  that can tolerate mutations and can potentially improve binding is important for designing targeted mutagenesis experiments. Instead of agnostically scanning single point and (a few) double mutations over all sites, these predictions can inform experiments to preferentially scan combinations of viable mutations on a smaller set of candidate sites. In previous work, evolutionary information was used to design such targeted mutagenesis for the HA and NA proteins of influenza~\cite{Wang2021-oh,Wu2020-zz}. A principled structure-based model could substantially improve the design of these experiments.

\subsection*{Discussion}
The success of AlphaFold has demonstrated the power of machine learning in predicting protein structure from sequence~\cite{Jumper2021-hl}. The challenge now is to leverage the experimentally and computationally determined protein structures to better understand and predict protein function. Our H-CNN model is a computationally powerful method to represent protein tertiary structures, and characterizes local biophysical interactions in protein micro-environments. Our model is physically motivated in that it respects rotational symmetry of protein structure data, allowing for significantly faster training time  compared to previous approaches~\cite{Torng2017-aq,Shroff2020-bu}. 

Similar to recent language models, H-CNN also demonstrates strong cross-task generalization by predicting quantitative effects of amino acid substitutions on function (i.e., zero-shot predictions), including protein stability or binding of protein complexes. Generally, massive language models trained on large and diverse protein sequence databases are shown to generalize well to predict mutational effects in proteins without any supervision~\cite{Alley2019-pp,Rao2019-dv,Rao2021-bv, Meier2021-sz, Rives2021-sr, Meier2021-sz}. State-of-the-art methods include ESM-1b for zero-shot predictions~\cite{Rives2021-sr} and MSA transformers that use evolutionary information from MSAs of protein families to predict the effect of mutations~\cite{Rao2021-bv}. The benchmark for these methods is the large set of DMS experiments, for which most zero-shot sequence-based predictions show an average accuracy of about 50\% in  predicting the rank order of the mutational effects~\cite{Meier2021-sz}. Our structure-based H-CNN method shows a comparable accuracy in predicting the mutational effect in DMS experiments of  the RBD protein in SARS-CoV-2, yet with much fewer parameters;  a more systematic analysis would be necessary to compare  these different approaches. Nonetheless, it would be interesting to see how the features extracted by H-CNN can complement the sequence-based language models to potentially improve zero-shot predictions for mutational effects in proteins.

Recent work has shown that combining structural data with evolutionary information from MSAs in deep learning models can be powerful in predicting mutational effects in proteins~\cite{Behbahani2022-zr}.  We have shown that H-CNN recapitulates the functional information reflected in  evolutionary data, further reinforcing the idea that physically guided structure-based machine learning models could be sufficient in predicting protein function, without a need for MSAs.  Importantly, our MSA-independent approach enables us to apply H-CNN to protein structures with no available homologs, including the {\em de novo} protein structures. 

The  H-CNN learned representations of amino acid neighborhoods could be used as input to a supervised  algorithm to learn a more accurate model for mutational effects in proteins; a similar approach has been used to model the stability effect of mutations in ref.~\cite{Blaabjerg2022-ok}. Moreover, the all-atom representation of protein structures used to train H-CNN allows for generalizability, e.g. using the inferred model to analyze non-amino acid molecules or extending the model and accommodate other elements to study protein-drug or protein-DNA interactions. 

Solving the inverse protein folding problem by designing a sequence that folds into a desired structure is a key step in protein design. Recent deep learning methods, including protein MPNN~\cite{Dauparas2022-fu} and transformer-based ESM-IF1~\cite{hsu_learning_2022,lin2022language}, have shown promise in designing viable sequences with a desired fold  for {\em de novo} proteins. H-CNN's ability to learn an effective potential in protein micro-environments merits investigation as to whether similar techniques can be used to solve the inverse folding problem for {\em de novo} proteins.

The learned  representation of protein micro-environments with H-CNN enables us to characterize the preferences of different amino acid compositions in a structural  neighborhood. Additionally, these rotationally equivariant representations could be used as building blocks of larger protein structure units, e.g. to characterize how  different molecular features on a protein surface could determine its interactions with other proteins.  A study in this direction could shed light on the structure-to-function map of the protein universe.

\section{Acknowledgment}
This work has been supported by the National Institutes of Health MIRA award (R35 GM142795), the CAREER award from the National Science Foundation (grant No: 2045054), the Royalty Research Fund from the University of Washington  (no. A153352), the Microsoft Azure award from the eScience institute at the University of Washington. This work is also supported, in part, through the Department of Physics, and the College of Arts and Sciences at the University of Washington.

%\bibliographystyle{plos2015}
%\bibliography{bib_armita,SI_bib}

\end{document}

\noindent{\bf \Large 
Supplementary Information\\}
\vspace{-1ex}

\noindent {\bf \large Learning the shape of  protein micro-environments  with\\  a holographic convolutional neural network}\\
\vspace{-2ex}

\noindent Michael N. Pun, Andrew Ivanov, Quinn Bellamy, Zachary Montague, Colin
LaMont, Philip Bradley, Jakub Otwinowski, and Armita Nourmohammad\\\\

 \tableofcontents
 \vspace{1cm}

%\input{sections_SI/data_details}
%

%!TEX root =  SI.tex
\section{Data pre-processing and code availability}

\noindent {\bf Code and data availability.} All codes and references to data are available on github through:\\ \href{https://github.com/StatPhysBio/protein_holography}{https://github.com/StatPhysBio/protein\_holography}. \\

\noindent{\bf Processing of the protein structure data.} We minimally process the protein structures  from the Protein Data Bank (PDB)~\cite{berman_protein_2000}.  We use PyRosetta~\cite{Chaudhury2010-mo} to infer coordinates of hydrogen atoms in the protein structure. We also use PyRosetta to assign solvent accessible surface area (SASA) and partial charge to all atoms.\\

\noindent{\bf Constructing the training, the validation, and the test sets for protein neighborhood classifications.} We classified amino acid neighborhoods extracted from the tertiary protein structures from the PDB. We used ProteinNet’s splittings for training and validation sets of the CASP12 competition to avoid redundancy due to similarities in homologous protein domains~\cite{alquraishi_proteinnet_2019}. Since PDB identifiers in ProteinNet were only provided for the training and the validation sets, we used ProteinNet’s training set as the base of both our training and validation and ProteinNet’s validation set as our testing set. Specifically, to define our training and validation sets, we make a [80\%,~20\%] split in the ProteinNet’s training data  of proteins with X-ray crystallography structures of 2.5\AA\, resolution or better. Furthermore, in anticipation of validating our model on experimental stabilities of T4 lysozyme and SARS-CoV2 receptor-binding domain (RBD), we removed all structures that belonged to the same UniProt family as the domains from each of these proteins. We used all of ProteinNet's validation structures as our testing set. 

This splitting resulted in 10,957 training structures, 2,730 validation structures, and 212 testing structures, each of which contained 2,810,503, 682,689, and 4,472 neighborhoods, respectively.\\

\noindent{\bf Data processing to assess the stability effect of mutations in the T4 lysozyme protein.} Predictions for the stability effect of mutations in the T4 lysozyme protein were made using the PDB structure 2LZM~\cite{weaver_structure_1987} as the wild-type. The  PDB identifiers for structures of all the T4 lysozyme variants used in our analyses (shown in Figs.~4~and~\ref{Fig:SI_T4}) along with the reported $\Delta\Delta G$ and the pH values of the stability measurements are reported in Table~\ref{tab:T4Lys}. 

Experimental measurements of $\Delta\Delta G$ values in variants for which we do not have a  matching protein structure (shown in Fig.~\ref{Fig:SI_T4_AF}) are taken from the dataset FireProtDB~\cite{stourac_fireprotdb_2021}. 

The AlphaFold prediction of the T4 lysozyme protein structure, used in Fig.~4C and~\ref{Fig:SI_T4_AF},  was performed using the AlphaFold GoogleColab notebook~\cite{Jumper2021-hl}.

Relaxation of mutant T4 lysozyme structures for the {\em in silico} model in Figs.~4B,C~and~\ref{Fig:SI_T4},  was done using PyRosetta \texttt{FastRelax} with the score function \texttt{ref2015\_cart}~\cite{Chaudhury2010-mo}. Backbone flexible positions were restricted to the mutated residue and the neighboring residues  in the sequence. Sidechain flexible positions were restricted to the neighbors with alpha carbons within 10 \AA\, of the mutated residue's alpha carbon.\\

\noindent{\bf Data processing to assess the fitness effect of mutations in the RBD of  SARS-CoV-2.} Deep mutational scanning experiments  from ref.~\cite{Starr2022-lp} were used to obtain the SARS-CoV-2 RBD expression and binding to the ACE2 receptor. Predictions of the RBD-ACE2 binding strength, shown in Fig.~5B,  were made using the  co-crystallized protein structure with the PDB identifier 6M0J~\cite{lan_structure_2020}. The predictions for the RBD stability, shown in Fig.~5A,  were made using the 6M0J structure with the ACE2 chain computationally removed. This computational removal was performed by selecting the RBD structure in PyMol~\cite{schrodinger_llc_pymol_2015} and exporting a PDB file from said selection. \\

\noindent{\bf Data processing to assess the effect of shearing in Protein G.} To analyze the effect of shearing  in protein G, shown in Figs.~3 and~\ref{Fig:SI_shear}, we used the native structure of protein G with the PDB identifier 1PGA~\cite{gallagher_two_1994}. Sheared structures were produced with PyRosetta by manually editing the backbone angles $\phi$ and $\psi$ of each residue (Fig.~3A) via  \texttt{Pose.set\_phi} and \texttt{Pose.set\_psi}~\cite{Chaudhury2010-mo}.

\section{Rotationally equivariant structure-to-function map for proteins}

The tertiary protein structure is a result of a folding process whereby amino acid sidechains interact with each other and the backbone structure to form the three-dimensional (3D) shape of the protein. The favorability of a certain amino acid occurring at a given site in a protein is determined by numerous factors: the presence of this amino acid during translation, the ability of the polypeptide chain to fold with that amino acid present, the amino acid's interactions with the local structural surroundings of the protein, and the amino acid's interactions with the surrounding environment. Our work focuses on revealing the third factor's contribution. Specifically, we aim to learn how a micro-environment in a protein's tertiary structure (i.e., atoms in a given structural neighborhood) constrain the  usage of different amino acids in the neighborhood. 

We define a structural neighborhood surrounding a focal amino acid by the coordinates (and identities) of all the $N$ atoms from the surrounding amino acids in the structure, within a given radius of the focal residue. Mathematically, we want to find a map $f: \mathbb{R}^{N \times 3} \rightarrow [{\mathbb{R}^+_{\leq 1}}]^{19}$  between the 3D coordinates of the $N$ neighboring atoms and the vector of probabilities $p\in [{\mathbb{R}^+_{\leq 1}}]^{19} $ associated with preferences for different types of amino acids that can be placed at the center of the neighborhood; note that the amino acid probability vector has 19 independent entries due to normalization. 

\begin{figure*}[p]
\centering\includegraphics[width=0.8\textwidth]{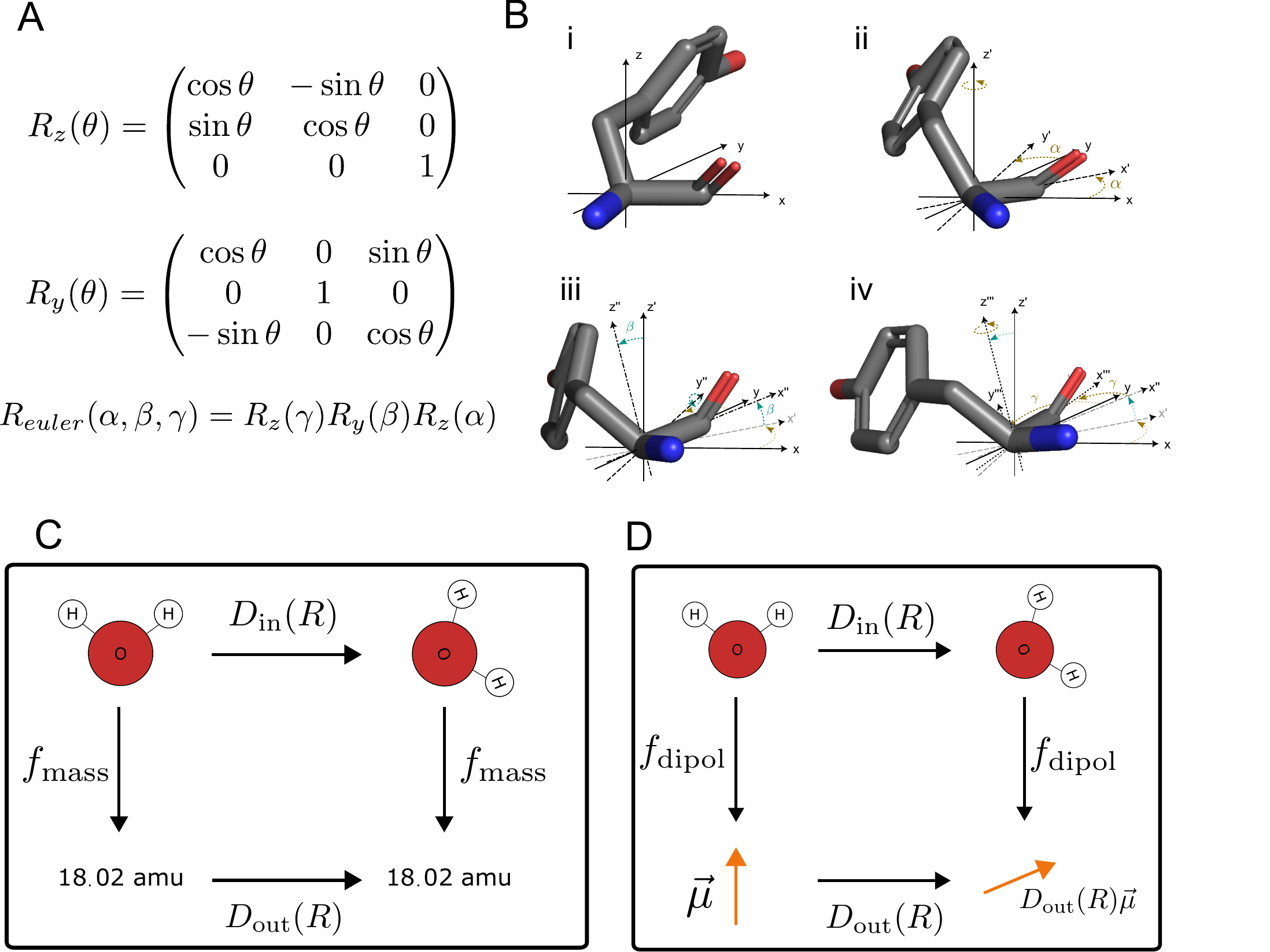}
\caption{
{\bf Parameterization of rotations and rotational equivariance.} 
We use x-y-z Euler angles to parameterize rotations in 3D. This parametrization {\bf (A)} can be  written in a matrix form, and {\bf (B)} it can be visualized  on a 3D object. 
Starting from a given orientation in (B.i), the object's rotation can be  broken down into three steps: First, rotation around the z-axis of angle $\alpha$~(B.ii). Second, rotation around the new y-axis by angle $\beta$~(B.iii). And finally, rotation around the new z-axis by angle $\gamma$~(B.iv). A map from the  geometric descriptions of an object to physical observables can be  invariant or equivariant under rotations. {\bf (C)} A rotationally invariant map, such as $f_{\rm mass}$ which gives the mass of a molecule, remains unchanged under the molecule's rotation. {\bf (D)} A rotationally equivariant map, such as $f_\text{dipol}$ which gives the dipole moment of a molecule, transforms  in a well-defined way with the molecule's rotation, denoted by the operator $D_{\rm in}(R)$. Specifically, in this case, the transformation operator for the dipole moment under rotation  $D_{\rm out}(R)$ acts linearly on the dipole moment in the original frame and is uniquely determined by the parameters of the rotation $R$.
\label{Fig:SI_euler}}
\end{figure*}

Assuming that the protein structure is a crystal (at least on average), the local interactions between the amino acid at the  focal site of interest  and the surrounding ones in the  protein structure have rotational symmetry, in that they only depend on the pairwise distances of atoms and not on their absolute orientation in space. This symmetry implies that the interactions are unchanged under global rotations since a global rotation preserves pairwise distances and relative orientations. Therefore, any model of amino acid preferences based on local structural information  should respect 3D rotational symmetry.  Thus, our goal becomes learning a map $f$, as described above, that is also invariant to global rotations acting on the inputs. A rotation acting on any Cartesian vector can be expressed as an Euler angle matrix $R_\euler$ (Fig.~\ref{Fig:SI_euler}). Since our inputs $\x$ for  coordinates of the atoms in a structural neighborhood  are Cartesian vectors $\x \in \mathcal{R}^{N\times 3}$, a global rotation  $R$ would act on these vectors via a Euler angle matrix, which we represent by $D_\cart(R)$. Therefore,  the rotationally invariant map should have the property $f(D_\cart(R) \x) = f(\x)$. 

In the rest of this section, we describe the mathematical foundation to construct 3D rotationally equivariant models for  protein structure data.
In the following section (Section~\ref{equivariance math}), we specifically  focus on our approach in encoding the structural data and training neural networks to learn  3D rotationally equivariant models for protein neighborhoods. The methodology discussed  in Sections~\ref{sec:equivariance encoding},~\ref{sec:equivariance processing} is self-contained, and if a reader wishes they can choose to skip Section~\ref{equivariance math} on the mathematical foundation of our approach.

\subsection{Mathematical foundation for 3D rotationally equivariant transformations}
\label{equivariance math}

The simplest approach to  train a neural network model that is invariant to  global rotations of proteins is to use invariant descriptors of the tertiary structure as inputs to the neural network. For example, one can  use dot products between all Cartesian vectors describing the  coordinates of  atoms within a structural neighborhood. However, recent work has shown that neural networks that are allowed to work with full geometric data as opposed to simple invariants have higher expressivity and perform better on learning tasks~\cite{batzner_e3-equivariant_2022}.

Networks that build more complex invariants rely on representation theory and the idea of {\em equivariance} to build  more expressive models. Equivariance is a generalization of invariance. Rather than requiring no change when inputs are transformed, equivariance allows the outputs of a map to change but requires them to change in a well-defined way. Under a transformation of the inputs corresponding to a rotation $R$, an equivariant map $f$ satisfies the relation 
\EQ
	f(D_\inp(R) x) = D_\out(R) f(x)
\EE
where $D_\inp(R)$ is the operator corresponding to the rotation in the input space, and $D_\out(R)$ is the operator corresponding to the rotation in the output space. Note that if $D_\out(R)$ is the identity matrix for all rotations, then the map is invariant.

The key to building an equivariant map is to know the operator $D_\out$ for the transformation of interest. The prevailing method in the field of equivariant machine learning is to modularly build a neural network out of equivariant units where each unit has an input and output space where these operators are known. The input and output space are often the same and tend to be the Fourier space defined by the symmetry that is respected.

Rotations in three dimensions are non-commutative which means that the output of two consecutive rotations can depend on the order by which they act on the input vector. As a result, the Fourier space for rotations is more complicated thaN that of a commutative transformation like translations.  Here, we will first use translations as an illustrative example for equivariant operations in Fourier space and then generalize to rotations.\\ 

\noindent {\bf Translational equivariance.} The Fourier transform of a signal on a real line $f(x)$ can be  expressed as $\hat{F}(k) = \int_{-\infty}^\infty e^{-ikx} f(x) dx$. We define a translated signal $f_{a}(x) = f(x - a)$ which is the original signal translated by a distance $a$. The Fourier transform of this translated signal follows
\EQA
\nonumber	\hat{F}_{a}(k) & = &\int_{-\infty}^\infty e^{-ikx} f_a(x) dx\\
\nonumber	 &=& \int_{-\infty}^\infty e^{-ikx} f(x-a) dx\\
\nonumber	& = & \int_{-\infty}^\infty e^{-ik(x'+a)} f(x') dx'\\
\nonumber	& =& e^{-ika} \hat{F}(k)\\
\EEA
Thus,  the Fourier transform of a translated signal transforms as $\hat{F}_a \overset{a}{\longrightarrow} e^{-ika}\hat{F}(k)$, implying that the Fourier transform is  equivariant under translation with an output operator that is simply a phase shift $D_\out (a) = e^{-ika}$. These operators $D_\out(a)$ form a representation of the group of 1D translations~\cite{tung_group_1985}.

In three dimensions, translations act on vectors via $\textbf{x} \rightarrow \textbf{x} + \textbf{a}$ and the representation is given by the operators operator $D_\out (\a) =  e^{-i\textbf{k}\cdot \bf a} $. \\

\noindent {\bf Rotational equivariance.} 
Similar to  translation, we can use Fourier space to define an equivariant  transformation for rotations. We consider rotations about a given reference point that  defines an origin for a spherical coordinate system in which points in 3D can be  expressed  by the resulting spherical coordinates $(r,\theta,\phi)$. Since the radius $r$ (i.e., the distance of a point from to the reference) does not change under rotations about the origin, we will ignore the radial component for now and consider a signal over the sphere of radius $r$, $f(\theta,\phi):~{S}^2(r) \rightarrow \mathbb{R}$. The Fourier transform $\hat F$ of the signal on the sphere follows, 
\EQ
\hat{F}_{\ell m}=\int_0^{2\pi}  \int_0^\pi f(\theta,\phi) Y_{\ell m} (\theta,\phi) \sin \theta \,\d\theta\,\d\phi
\EE
where $Y_{\ell m}(\theta,\phi)$ is the spherical harmonic of degree $\ell$ and order $m$ defined as
\EQ
	Y_{\ell m}(\theta,\phi) = \sqrt{\frac{2n+1}{4\pi}\frac{(n-m)!}{(n+m)!}}e^{im\phi}P^m_\ell(\cos\theta)
	\label{eq:YLM}
\EE
where $\ell$ is a non-negative integer ($0\leq \ell$), and $m$ is an integer within the interval $-\ell\leq m \leq \ell$. $P^m_\ell(\cos \theta)$ is the Legendre polynomial of degree $\ell$ and order $m$, which, together with the complex exponential $e^{im\phi}$, define sinusoidal functions over the angles $\theta$ and $\phi$. In quantum mechanics, spherical harmonics are used to  represent  the orbital angular momenta, e.g. for an electron in a hydrogen atom. In this context, the degree $\ell$ relates to the eigenvalue of the square of the angular momentum, and the order $m$ is the eigenvalue of the angular momentum  about the azimuthal axis.

The operators that describe how spherical harmonics transform under rotations are called the Wigner D-matrices, denoted by  $D^{\ell}_{mm'}(R)$~\cite{tung_group_1985}; this operator is a matrix  because the rotation group in 3D is not commutative. Under a rotation $R$, spherical harmonics transform as
\EQ
	Y_{\ell m}(\theta,\phi) \overset{R}{\longrightarrow}\sum_{m' = -\ell}^\ell D^{\ell}_{m'm}(R) Y_{\ell m'}(\theta,\phi)
	\label{eq:YLMWignerD}
\EE

Since spherical harmonics transform in this well-defined way, we can now examine how a rotation acts on the Fourier transform of a signal over the sphere. Suppose we have a signal $f({\textbf{n}})$ defined on the elements ${\bf n}$ of the spherical shell $S^2$ (i.e., the set of angular coordinates for points on the sphere). The Fourier coefficients associated with the rotated signal  $f_R({\textbf{n}}) = f(R{\textbf{n}})$ follow, 
\EQA
\nonumber 	\hat{F}^R_{\ell m} = \int Y_{\ell m}({\textbf{n}}) f_R({\textbf{n}}) \d\Omega &=& \int Y_{\ell m}({\textbf{n}}) f(R{\textbf{n}}) \d\Omega\\ 
\nonumber	& = &\int Y_{\ell m}(-R{\textbf{n}}') f({\textbf{n}}') \d\Omega'\\ 
\nonumber	& = &\int \sum_{m' = -\ell}^\ell D^\ell_{m'm}(-R) Y_{\ell m}({\textbf{n}}') f({\textbf{n}}') \d\Omega'\\ 
\nonumber	& = &\sum_{m' = -\ell}^\ell D^\ell_{m'm}(-R) \int Y_{\ell m}({\textbf{n}}') f({\textbf{n}}') \d\Omega'\\
\nonumber & = &\sum_{m' = -\ell}^\ell D_{m'm}^\ell(-R) \hat{F}_{lm}\\
\EEA
where $\d\Omega =  \sin^2 \theta \,\d\theta\,\d\phi $ is the angular differential in the spherical coordinate system. Under a rotation $R$ about the origin, the spherical Fourier coefficients of a real-space signal transform according to $\hat{F}_{\ell m} \overset{R}{\longrightarrow}\sum_{m' = -\ell}^\ell D^\ell_{m'm}(-R)\hat{F}_{\ell m'}$, which is simply a matrix product.  

It should be noted that the Wigner D-matrices form an irreducible representation of SO(3), which is the group of all rotations around a fixed origin in three-dimensional Euclidean space.
Therefore, it provides a natural basis to expand functions in SO(3) while respecting the rotational symmetry.\\

\noindent {\bf Nonlinear transforms respecting rotational equivariance.} One key feature of neural networks is applying nonlinear activations,  which enable a network to approximately model complex and non-linear phenomena. Commonly used nonlinearities include reLU,  tanh, and softmax functions. However, these conventional non-linearities can break rotational equivariance in the Fourier space. To construct expressive rotationally equivariant neural networks we can use the Clebsch-Gordan tensor product, which is the natural nonlinear (or bilinear in the case of using two sets of Fourier coefficients) operation in the space of spherical harmonics~\cite{tung_group_1985}.

Given two spherical tensors $\hat{F}_{\ell_1}$ and $\hat{G}_{\ell_2}$, we can take a product between all of their components $\hat{F}_{\ell_1m_1}\hat{G}_{\ell_2m_2}$, which would allow us to express nonlinearities. Although these products do not behave well under rotations, linear combinations of them do transform in well-defined ways. The Clebsch-Gordan coefficients are the coefficients that decompose these products back into the space of spherical harmonics,  where Wigner D-matrices  define the rotation operations. Specifically, the product of spherical tensors $\hat{F}_{\ell_1}$ and $\hat{G}_{\ell_2}$ will yield spherical tensors  of degree $L$ such that $|\ell_1 - \ell_2| < L <\ell_1 + \ell_2$ in the following way,
\EQ
	\hat{H}_{LM} = \sum_{m_1 = -\ell_1}^{\ell_1} \sum_{m_2 = -\ell_2}^{\ell_2}  \CG{\ell_1}{m_1}{\ell_2}{m_2}{L}{M}\hat{F}_{\ell_1m_1}\hat{G}_{\ell_2m_2}
\EE
where $ \CG{\ell_1}{m_1}{\ell_2}{m_2}{L}{M}$ is a Clebsch-Gordan coefficient and can be precomputed for all degrees of spherical tensors~\cite{tung_group_1985}. Similar to spherical harmonics, Clebsch-Gordan products also appear in quantum mechanics, and they are used to express couplings between angular momenta. In following with recent work on group-equivariant machine learning~\cite{Kondor2018-tr,Thomas2018-uz,batzner_e3-equivariant_2022,musaelian_learning_2022}, we will use Clebsch-Gordan products to express nonlinearities in 3D rotationally equivariant neural networks for protein structures.

\subsection{Rotationally equivariant encoding of protein  micro-environments as spherical holograms}
\label{sec:equivariance encoding}

We define an  amino acid's micro-environment as all the atoms associated with the surrounding amino acids  within a 10\AA\, of the central residue’s $\alpha$-carbon. We use the position of the central residue’s $\alpha$-carbon as the reference point of the neighborhood $\mathcal{N}$. Each atom $i$ within this neighborhood, located at position $\textbf{r}_i$ with respect to the reference point, has three attributes associated with it: (i) the element type of the atom,  $e_i \in \{C,N,O,S,H\}$, (ii) the partial charge of the atom $Q_i$,  and  (iii) the SASA $A_i$ of the atom. With these characteristics we define a feature vector $v_i^{c}$ where $c$ indexes the input channels $\{C, N, O, S, H, \text{charge},  \text{SASA}\}$. This vector takes on values given by 
\begin{align*}
	v_i^c = \begin{cases}
		\delta_{e_i,c} &\text{ for } c \in \{C, N, O, S, H\}\\
		Q_i &\text{ for } c = \text{charge}\\
		A_i &\text{ for } c = \text{SASA}
	\end{cases}
\end{align*}
where $\delta_{ij}$ is a Kronecker delta function, and takes the value 1 if $i=j$ and 0 otherwise. Thus, for $c \in \{C, N, O, S, H\}$ the feature vector  $v_i^c$ takes value 1 if atom $i$ is of type $c$ and 0 otherwise. $Q_i$ and $A_i$ denote the partial charge and the SASA (in units of ${\rm \AA}^2$) of atom $i$, respectively. These quantities are computed by  PyRosetta for each protein structure~\cite{Chaudhury2010-mo}.

Each channel defines a point cloud with the attributed values stored at the coordinates of the corresponding atoms.  We express the point cloud of each channel in a  spherical coordinate system with the origin set at the alpha carbon of the central amino acid in the neighborhood. We then define an atomic density as a sum of Dirac delta distributions parameterized by each atomic coordinate in the neighborhood $\rho^c(\textbf{r}) = \sum_{i\in \mathcal{N}} v^c_i\delta^{(3)}(\textbf{r} - \textbf{r}_i)$. Here $\delta^{(3)}(\bf x)$ denotes a normalized Dirac delta probability density in 3D, which is zero everywhere except at the origin $\bf x=0$, where it is infinite.  We use 3D Zernike transforms to encode the point clouds associated with each channel in a 3D rotationally  equivariant fashion in the spherical Fourier space:
\EQ 
\hat{Z}^c_{n\ell m} = \int \rho^c({\bf r}) Y_{\ell m}(\theta,\phi) R _{n \ell}({\bf r})  \,\d \Omega.
\label{eq:Zernike}
\EE
where  $\d\Omega$ is the angular differential in the spherical coordinate system, $Y_{\ell m}(\theta,\phi) $ is the spherical harmonics (eq.~\ref{eq:YLM}), and $R _{n \ell}({\bf r})$ is the radial function of the 3D Zernike transform, which can be expressed as,
\EQ
R _{n \ell}({\bf r}) =  (-1) ^{\frac{n-\ell}{2}} \sqrt{2n +3} \,
{ \frac{n+\ell+3}{2} -1 \choose \frac{n-\ell}{2}} |{\bf r}|^{\ell}\, {}_{2}F_{1}\left ( - \frac{n-1}{2},\frac{n+\ell +3}{2}; \ell + \frac{3}{2}; |{\bf r}|^2\right)
\EE
where ${}_{2}F_1(\cdot)$ is the ordinary hypergeometric function.  The index $n\geq 0$ is a non-negative  integer, and the radial function $R_{n\ell}(\bf r)$ is nonzero only for even values of $n-\ell\geq 0$. Zernike polynomials form a complete orthonormal basis in 3D and, therefore, can be  used  to expand and retrieve any 3D shape, if large enough $\ell$ and $n$ coefficients are used. Approximations that restrict the series to finite $n$ and $\ell$ are often sufficient for shape retrieval, and hence, desirable algorithmically. Importantly, expansion of an input signal  by Zernike  polynomials (eq.~\ref{eq:Zernike}) is  rotationally equivariant since a rotation $R$ transforms the resulting coefficients through Wigner-D matrices as  $\hat{Z}_{n\ell m}^c \overset{R}{\longrightarrow} \sum_{m' =-\ell}^\ell D^\ell_{m'm}(-R)\hat{Z}_{n\ell m'}^{c}$ (eq.~\ref{eq:YLMWignerD}).  

Zernike projections in the spherical Fourier space can be thought of as a superposition of spherical holograms of an input point cloud, and thus, we term this operation the {\em holographic encoding} of protein micro-environments.

\subsection{Rotationally equivariant model of protein micro-environments with H-CNN}
\label{sec:equivariance processing}
We use the coefficients of the Zernike expansion $\hat{Z}_{nlm}^{c}$  in eq.~\ref{eq:Zernike} (i.e., the holographic encoding of the data) as inputs to a rotationally equivariant convolutional neural network to classify  amino acids based on their surrounding atomic micro-environments. We term this network holographic convolutional neural network (H-CNN).

The convolutional neural network that we use for our analysis is similar to the Clebsch-Gordan network (CGNet), described in ref.~\cite{Kondor2018-tr}. This network is built out of three rotationally equivariant modular units: (i) linear transformation of spherical harmonics, (ii) Clebsch-Gordan nonlinearity, and (iii) spherical batch normalization.

Below we will describe the computations done in each of these modular equivariant units. Without a loss of generality, we denote the operations in each unit of the network in terms of a general equivariant signal $\hat{F}^k_{lm}$ where $k$ is a catch-all index that represents all non-rotational indices (i.e., channel $c$ and index $n$ in eq.~\ref{eq:Zernike}), and $\ell$ and $m$ correspond to the angular indices (eq.~\ref{eq:YLM}). The output of each unit will be denoted by $\hat{G}^{k}_{lm}$.\\

\noindent {\bf Linear transformations in CG-nets.} Linearities are ubiquitous in neural networks and contribute partially to their powerful expressiveness. Our linearity takes the form
\begin{equation}
	\hat{G}^k_{\ell m} = \sum_{k'}W^{(\ell)}_{kk' } \hat{F}^{k'}_{\ell m}
	\label{eq:mixingLin}
\end{equation}
where $W^{(\ell)}_{kk'}$ denotes two dimensional matrices that mix different input channels $k'$ (e.g. combinations of radial index $n$ and channel $c$ in the input or general learned representations in intermediate layers) producing different output channels $k$. To preserve  rotational symmetry, we will impose the restriction of only combining information (i.e., forming linear combinations of inputs) that belongs to the same irreducible representations (irreps) of SO(3) in eq.~\ref{eq:mixingLin}. In other words, we mix only inputs associated with the same degree $\ell$ and order $m$  of spherical harmonics. These inputs are mixed across channels associated with different chemical or radial information in the first layer, or composite channels in the following layers. Generally, different weights are learned for each irrep $\ell$ but shared across coefficients of different order $m$.\\

\noindent {\bf  Clebsch-Gordan nonlinearity.} The last equivariant component of the network is a Clebsch-Gordan nonlinearity. Nonlinearities are  ubiquitous in neural networks. The Clebsch-Gordan nonlinearity is the simplest equivariance-respecting nonlinearity, and it is expressive enough for most learning tasks~\cite{Kondor2018-tr,musaelian_learning_2022}. This Clebsch-Gordan nonlinearity is simply a product of spherical tensors that is decomposed back into the spherical harmonic basis via the Clebsch-Gordan coefficients,

\begin{equation}\label{eqn:CG_product}
	\hat{H}^{K}_{LM} = \sum_{m_1 = -\ell_1}^{\ell_1} \sum_{m_2 = -\ell_2}^{\ell_2}  \CG{\ell_1}{m_1}{\ell_2}{m_2}{L}{M}\hat{F}^{k_1}_{\ell_1m_1}\hat{G}^{k_2}_{\ell_2m_2} 
\end{equation}
where $\CG{\ell_1}{m_1}{\ell_2}{m_2}{L}{M}$ is a Clebsch-Gordan coefficient  and $K = (k_1,k_2)$ is the new channel index.

Generally we have a choice for which combinations of values $k_1,k_2,\ell_1,$ and $\ell_2$ are used in this product. We focus on two choices for these indices. We define {\em fully connected} networks, for which $k_1$, $k_2$, $\ell_1$, and $\ell_2$ are allowed to take on all possible values. We also define {\em simply connected} networks, for which we impose the condition that $k_1 = k_2$ and $\ell_1 = \ell_2$. \\

The exact choice made in combining these indices affects the dimensionality of the network. The width of the  Clebsch-Gordan nonlinearity output   in a fully connected network with a maximum spherical degree  $L_{\max}$ and  a width  $d_h$ (i.e., $k \in \{1,...,d_h\}$) is given by
\begin{equation}\label{eqn:omega fc}
	\Omega_{\ell,L_{\max}}^{\rm full} = d_h^2 \bigg( \bigg[\frac{1}{4}(2 + \ell(5+\ell) - 2 \Big\lfloor \frac{1}{2}(\ell - 1) \Big\rfloor\bigg] + (\ell + 1)(L _{\max}- \ell)\bigg) + (\ell + 1)(L_{\max} - \ell)
\end{equation}
Meanwhile, for a simply connected network with the same $L_{\max}$ and $d_h$, the width of the  Clebsch-Gordan nonlinearity is 
\begin{equation}\label{eqn:omega sc}
	\Omega_{\ell,L_{\max}}^{\rm simple} = d_h (L_{\max} + 1 - \ell)
\end{equation}

\noindent{\bf Spherical batch normalization.} Batch normalization is a common feature in neural networks that allows for smooth training of the network. Since the output of our nonlinear Clebsch-Gordan product is not bounded, batch normalization becomes extremely important  to ensure that activations remain finite throughout training. We impose a batch normalization layer after each linear operation, producing normalized inputs to the nonlinear operation (Clebsch-Gordan product). We use the following batch normalization during training,
\begin{align}
	\hat{F}^{k}_{\ell m} = \frac{\hat{F}^{k}_{\ell m}}{\sqrt{\langle |\hat{F}^{k}_{\ell}|^2  \rangle_\text{batch}+ \epsilon}}
\end{align}
where  $|\hat{F}^{k}_{\ell}|^2 = \sum_{m=-\ell}^\ell \hat{F}^k_{\ell m}\hat{F}^{*k}_{\ell m}/(2\ell+1)$, $\phantom{0}^*$ denotes complex conjugation, $\langle\cdot \rangle_\text{batch}$ denotes averaging over a batch, and $\epsilon$ is a small number used to ensure numerical stability, which we set to $\epsilon = 10^{-3}$. During training the network  stores a moving average of the norm for each $\ell$ in each channel as
\begin{align}
	N^{\mu,i}_{\ell} = \xi \langle |\hat{F}^{k}_{\ell} | ^2 \rangle_\text{batch}+ (1 - \xi) N^{\mu,i-1}_{\ell}
\end{align}
where $\xi$ serves as the momentum of our moving average, set to $\xi = 0.99$. This moving average is then used as the norm during testing in the following way:
\begin{align}
	\hat{F}^{\mu}_{\ell m} = \frac{\hat{F}^{\mu}_{\ell m}}{\sqrt{N^{\mu,i}_{\ell} + \epsilon}}
\end{align}

The different batch normalization used for training and testing prevents the batching of the validation data to affect the evaluation of the model accuracy during testing.

\section{Network and training procedure for H-CNN}

\begin{figure*}[p]
\includegraphics[width=\textwidth]{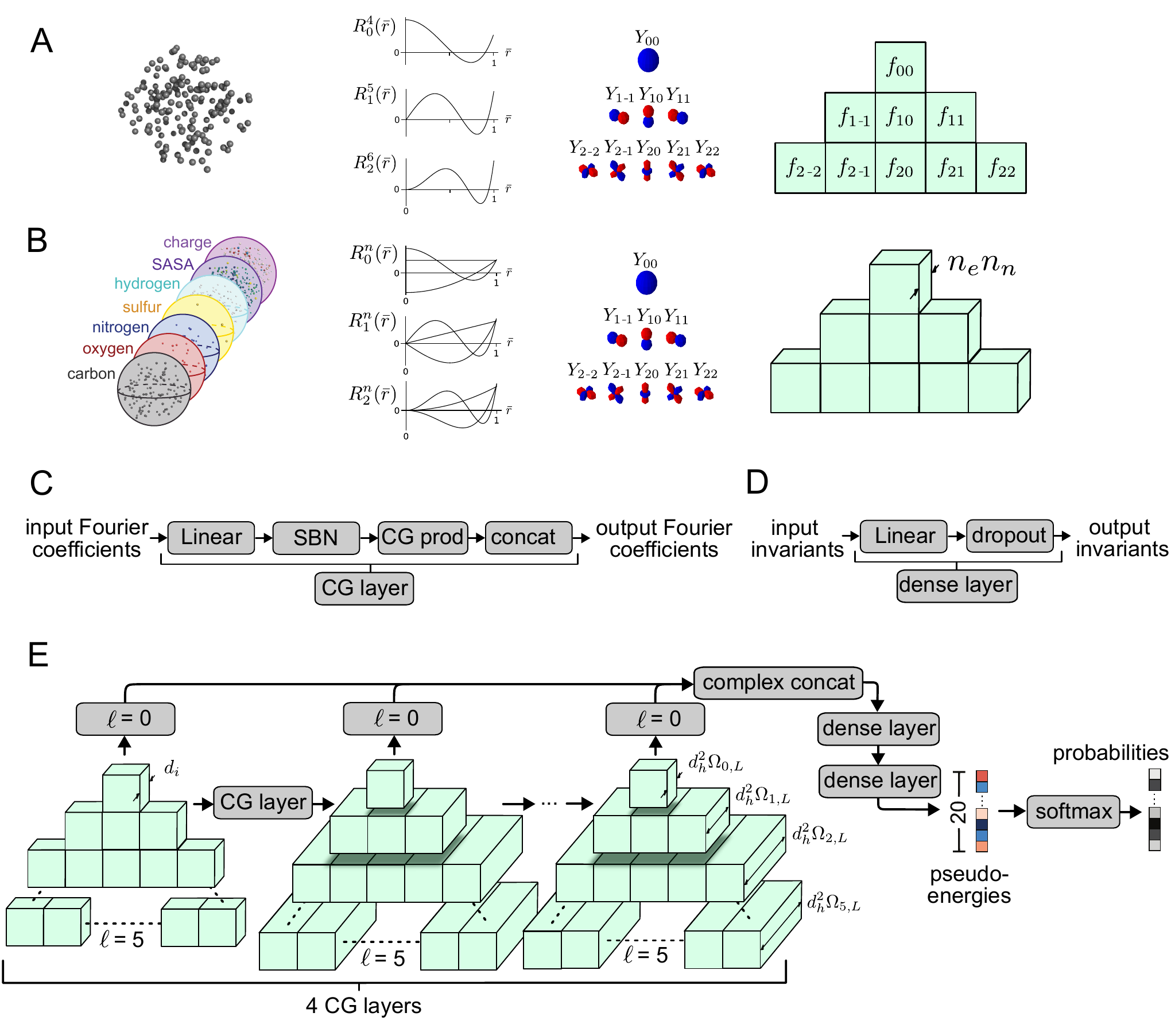}
\caption{
{\bf  Schematics of rotationally equivariant data encoding and network architecture in H-CNN.} 
{\bf (A)} Any point cloud (left) can be written in the equivariant spherical harmonic Fourier basis by integrating the point cloud density against the spherical harmonics along with a choice of a radial function (middle), which depends on the spherical harmonic degree $\ell$. The triangular structure (right) demonstrates the resulting Fourier coefficients $f_{\ell m}$, in which the rows reflect the different  degrees of spherical harmonics   $0\leq \ell $, which are non-negative integers. Individual cells within each row correspond to different orders $m$ of the spherical harmonics, which are integers bounded by $-\ell \leq m\leq \ell$. {\bf (B)} The Fourier transform becomes three dimensional when radial resolution is required thereby increasing the number of discrete $n$-values used $n_n$. The multiplicity of coefficients is also determined by the number of point clouds being transformed $n_\rho$ (e.g. the different atomic channels on the left). Since both the point clouds and the radial information are invariant under rotations, these two dimensions can be combined in index to give preference in organizing information according to the degree $\ell$ and order $m$ of the spherical harmonics. {\bf (C)} The Clebsch-Gordan block (CG block) is fully equivariant and is comprised of a linear layer, a spherical batch norm (SBN), Clebsch-Gordan product (CG prod), and a degree-wise concatenation (concat).
{\bf (D)} Throughout all layers of the network, all invariants ($\ell = 0$) are collected and fed into a series of dense layers which are simple linear combinations with training dropout. {\bf (E)} These two main components in (C,D) comprise the bulk of the entire network. First the input Fourier coefficients are processed through successive CG layers. Invariants are collected from the original input and from the output of every CG layer. The real and imaginary components of these invariants are split and concatenated (complex concat) and then fed into a series of dense layers. This schematic reflects the dimensions of the optimal model discovered in this work. Four CG layers were used. We note that the maximal width of the network as determined by the output of the nonlinearity depends on $\ell$ due to the selection rules of the CG prod. We denote this width $d^2_h \Omega_{\ell,L}$ where $L$ is the maximal degree $\ell$ used in the network and are given in~\cref{eqn:omega fc,eqn:omega sc}. Two dense layers were used in the optimal model. The dimensions of these layers are $(4852 \times 500)$ and $(500 \times 20)$. The output of these dense layers are termed pseudo-energies and act as logits in a softmax to estimate probabilities for each amino acid. We emphasize the only nonlinearities in the network are the Clebsch-Gordan products and the ultimate softmax. \label{Fig:SI_network}}
\end{figure*}

\noindent{\bf Network architecture.} A complete H-CNN features an equivariant unit of Clebsch-Gordan layer followed by an invariant unit of dense layers as detailed in Fig.~\ref{Fig:SI_network}. A Clebsch-Gordan layer is comprised of an equivariant linearity, a spherical batch norm, a Clebsch-Gordan product, and a concatenation. Invariants from the inputs as well as all outputs of the each CG layer are collected for use in the invariant dense layers (Figs.~1,~\ref{Fig:SI_network}). \\

\noindent{\bf Training procedure.} Networks were trained with all training data being read at run time using the keras \texttt{fit} function and a data generator made from a subclass of the \texttt{keras.utils.sequeunce} class~\cite{chollet_keras_2015}. Generally 40 workers were used for reading the data and a \texttt{max\_queue\_size} of 900 was found to be optimal  for read/evaluation balance. Approximately one tenth of the validation data (51,200 neighborhoods) was evaluated after an approximate one tenth of the training data (256,000 neighborhoods) was seen. This was implemented by using \texttt{tf.keras.Model.fit} arguments \texttt{steps\_per\_epoch=1000} and \texttt{validation\_steps=100}. Data was shuffled at the end of each evaluation interval. These evaluation intervals also served as checkpoints for both early stopping and weight saving. The best network was trained for 8.47 epochs in 4.54 hours on one NVIDIA A40 GPU hosted at the Hyak supercomputer at the University of Washington.

We used stochastic gradient descent and Adam optimizer~\cite{kingma_adam_2015} for training our networks. Adam generally showed better performance, and thus, all networks were trained with Adam during hyperparameter optimization. The main component of the loss was defined as the categorical cross entropy between the one-hot amino acid encoding and the softmax output of the network
\begin{equation}\label{eqn:val_loss}
	\mathcal{L} = -\sum_{i = 1}^N \log p_{\alpha_i}
\end{equation}
where $p_{\alpha_i}$ is the network-assigned probability of the true central amino acid $\alpha_i$ in neighborhood $i$.

Some hyperparameters were scheduled according to the behavior of the network's loss. The learning rate was put on a schedule of \texttt{ReduceLROnPLateau} to decrease by a factor of 0.1 whenever the validation loss did not improve over 10 epochs, with a minimum learning rate of $10^{-9}$.  Ultimately, the networks were trained with early stopping when the validation loss did not improve by a difference of 0.01 over the course of 20 epochs.\\

\begin{figure*}[p]
\includegraphics[width=6in]{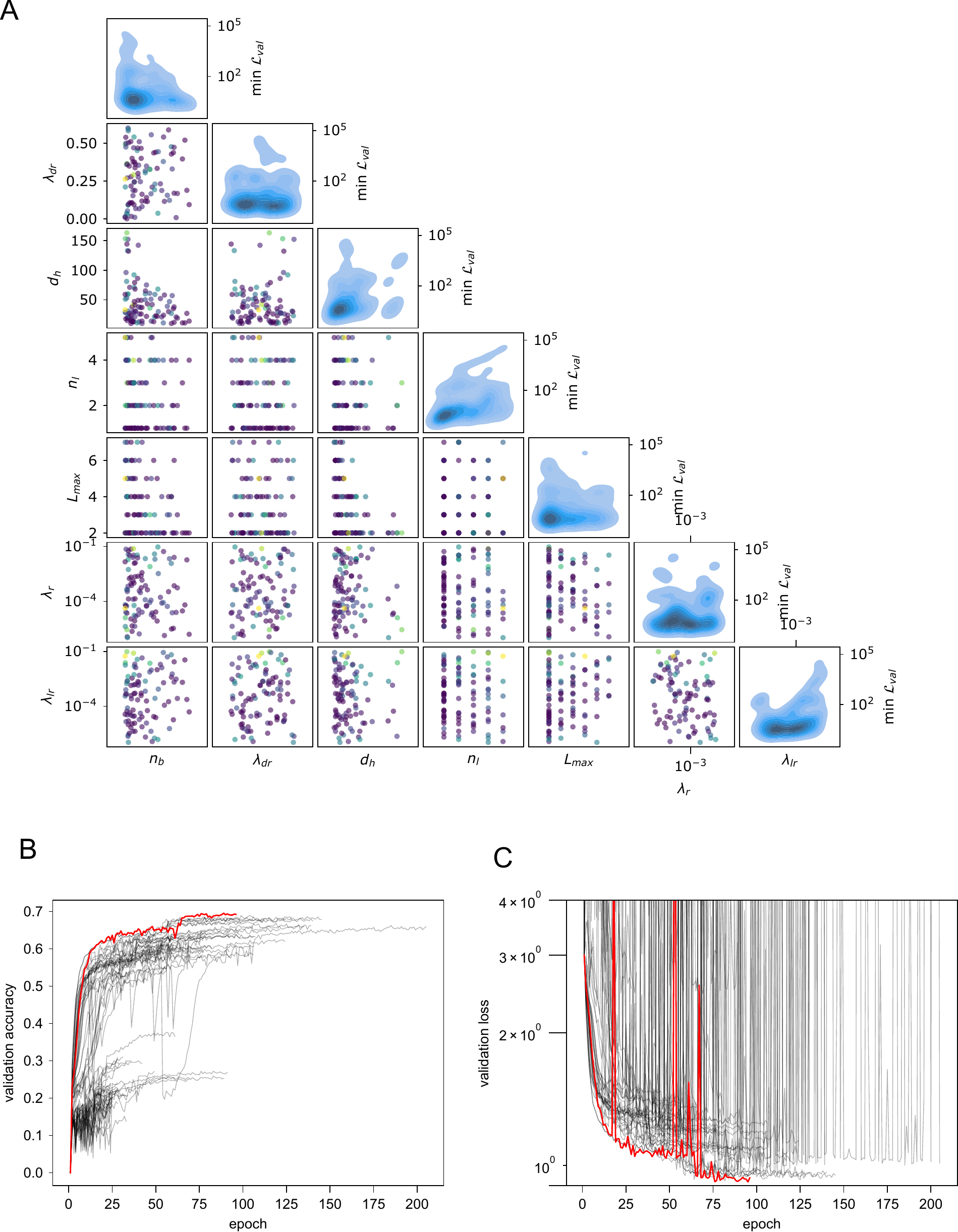}
\caption{
{\bf Hyperparameter optimization for H-CNN.} 
Hyperparameter optimization was performed in two steps for both the simply connected and the fully connected networks: First, hyperparameters were scanned across a wide regime of hyper-parameters via random sampling. From this scan, a narrow band of hyper-parameters were analyzed to determine the optimal network {\bf (A)}. Off the diagonal, each dot's color shows the validation loss seen during this first step of hyperparameter tuning for a network with the combination of hyperparameters shown on the two axes. On the diagonal, the density of networks is shown as a function of loss and the hyperparameter tested.  Validation accuracy {\bf (B)} and loss {\bf (C)} are shown as a function of training epochs for all networks trained in the second stage of hyperparameter tuning. Red lines show the network with the best validation loss that is used throughout this work.  
\label{Fig:SI_training}}
\end{figure*}

\noindent{\bf Hyperparameter optimization.}  Hyperparameter optimization was performed in two steps: First, we scanned a larger parameter regime and trained 500 networks to identify a reasonable subregion of the parameter space (Fig.~\ref{Fig:SI_training}A). This sampling revealed networks with parameter ranges listed in Table~\ref{table:hyper}. These parameters were then sampled more finely to get the best network models. The resulted training curves are shown in Fig.~\ref{Fig:SI_training}B,C.

It should be noted that we optimized over two classes of networks: (i) fully connected networks, and (ii) the {simply-connected} networks, as defined in Section~\ref{sec:equivariance processing}. 
In simply connected networks, the nonlinear Clebsch-Gordan products are restricted to the square of Fourier coefficients with the same $\ell$ and channel index, in contrast to the fully connected networks that use bilinear products between all sets of Fourier coefficients (see equation~\eqref{eqn:CG_product}). These restricted nonlinearities allowed us to use linear layers that had roughly one order of magnitude larger output dimensions in simply vs. fully connected networks (150 as opposed to 20); see Fig.~\ref{Fig:SI_training}E. We separately optimized the hyperparameters for these two network classes.  \\

\noindent{\bf Model performance.} For both the simply connected and the fully connected networks, we find the best hyperparameters that minimize the validation loss function in equation~\ref{eqn:val_loss}. The best simply-connected model had 62\% classification accuracy and a minimal validation loss of 1.2 while the best fully-connected model had 68\% classification accuracy and a minimal validation loss of 0.91. This difference in predictive power of the models appeared consistent across all training scenarios.

We compare  the accuracy and the training efficiency of our  model to the existing methods that classify amino acids based on all-atom representations of their local environments in Table~\ref{tab:performance}. For this comparison, we used the metric of testing accuracy of the amino acid class since this was more commonly reported in the literature. Specifically, we compare H-CNN with two 3D CNN methods that used conventional convolutional neural networks on voxelized protein structures~\cite{Torng2017-aq,Shroff2020-bu}. Notably, these 3D CNN methods require local alignment of neighborhoods to the central amino acid. Also, each of these methods uses a cube of side length 20\AA{}, while spheres of radius  10\AA{} are used as inputs to H-CNN. Our method thus sees approximately half ($\pi/6$) of the volume that these other models see. Overall, our model has a comparable accuracy to the best 3D CNN method~\cite{Shroff2020-bu}, but it is much more efficient in training. 

We also compare our model to other methods that prioritize rotational symmetry, i.e., the Spherical CNN introduced in ref.~\cite{boomsma_spherical_2017} and the 3D Steerable CNN from ref.~\cite{weiler_3d_2018}. Spherical CNN~\cite{boomsma_spherical_2017} is approximately rotationally invariant by performing 3D convolutions on  a discretized grid over the 3D sphere. 3D Steerable CNN~\cite{weiler_3d_2018} is a rotationally invariant method by applying spherical convolutions in a steerable basis. H-CNN performs better than both of these methods in the classification task with an order of magnitude fewer parameters (Table~\ref{tab:performance}). \\

\begin{figure*}[p]
\includegraphics[width=\textwidth]{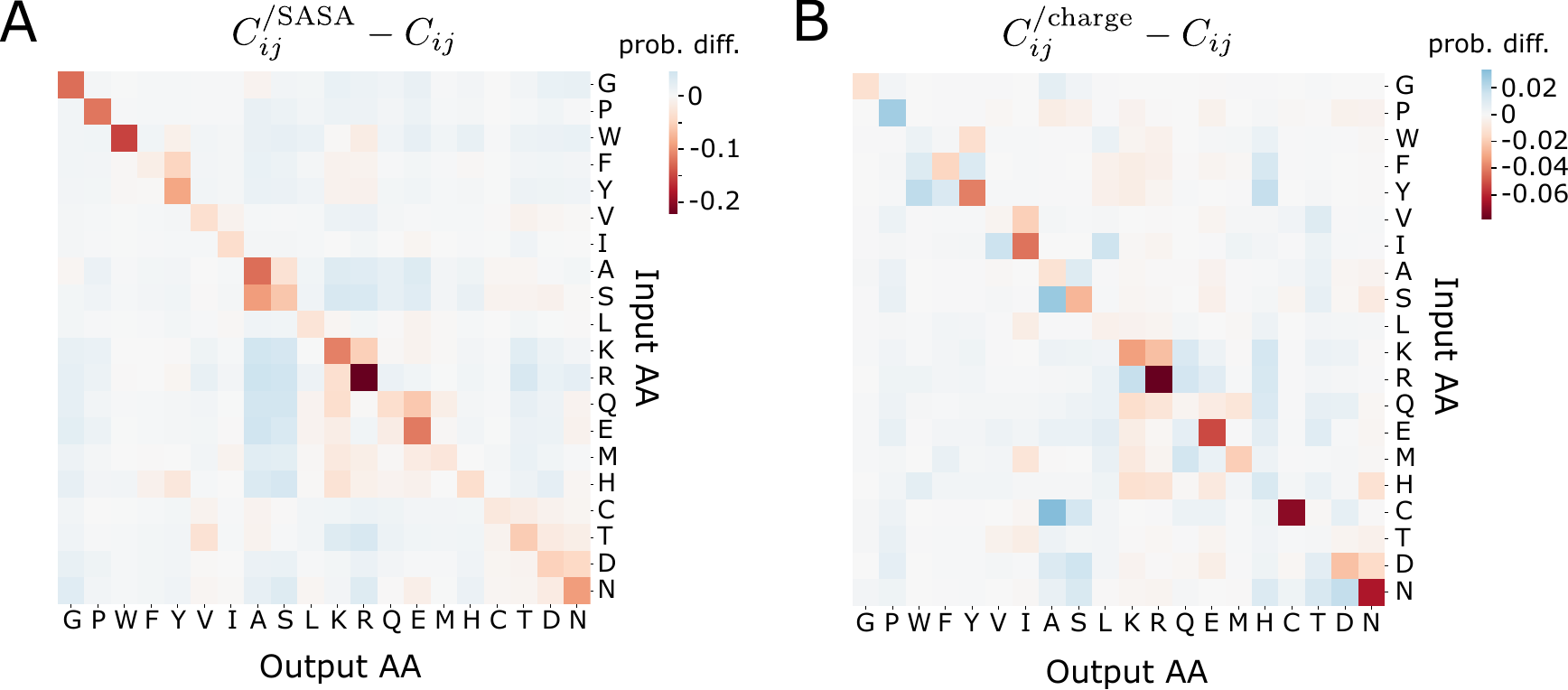}
\caption{
{\bf H-CNN performance  after  charge and SASA ablation.} 
The difference in the confusion matrix for networks with ablated {\bf (A)} SASA and {\bf (B)} charge with respect to the network with the full information (Fig.~2A) is shown. Negative (red) values demonstrate less probability assigned to the input/predicted pair when the input is ablated, while positive (blue) values show increased probability mass under ablation. The removal of SASA (A) appears to decrease the network's ability to predict all amino acids as seen on the diagonal. The effect is most pronounced for hydrophobic amino acids in the upper left, with some hydrophilic amino acids (R, K, E) also displaying strong effects. Interestingly, most of the probability mass is on average redistributed to small amino acids A and S. When charge is removed (B), the network demonstrates worse predictions on charged and polar amino acids most notably R, C, N, and E. The overall network's accuracy after removing SASA and Charge are 57\% and 56\%, respectively. 
\label{Fig:SI_cmat}}
\end{figure*}

\noindent{\bf Alternative models.} We studied the effect of our post-processing of atomic neighborhoods by performing ablation studies on  charge and SASA. For each ablation, we reran the second stage of our hyperparameter optimization and only considered fully connected networks. The overall accuracy from removing SASA was 57\% while the overall accuracy from removing charge was 56\%, in contrast to the 68\% accuracy of the complete model shown in Fig.~2. A similar 10\% drop in accuracy associated with SASA and charge was previously reported in ref.~\cite{Shroff2020-bu}.  Fig.~\ref{Fig:SI_cmat} shows the confusion matrices for each model compared to the best model's confusion matrix. 

\section{Comparing H-CNN with evolutionary models for amino acid preferences} 
As shown in the main text, the interchangeability of amino acids predicted by H-CNN is consistent with the evolutionary substitution patterns reflected by the BLOSUM62 matrix (Fig.~2D). In addition to this coarse-grained measure, we also assessed the consistency of substitutions at the single-site level between the H-CNN predictions and evolutionary data. Evolutionary variation at a given site in a  protein reflects both the substitutability of amino acids at  that position (e.g. due to their common physico-chemical properties) and the epistatic interactions with the  other covarying residues in the protein. 

To isolate the site-specific effects of substitutions, we used the software EV-Couplings~\cite{sheridan_evfoldorg_2015} to infer a model for amino acid preferences (and their couplings), using amino acid covariations in multiple sequence alignments (MSAs) of protein families. Specifically, EV-Couplings infers a maximum likelihood Potts model for interactions between residues in a protein sequence~\cite{weigt_identification_2009}. From this model, we can evaluate the  relative preference for occurrence of different amino acids  at a given position in a protein sequence. 

We used EV-Couplings to infer models of amino acid preferences for all of the protein domains in our test set. For each protein chain in our test data, we used EV-Couplings to compute the MSA for the protein associated with UniProt ID listed under the associated PDB entry.  The MSA is computed using UniRef90~\cite{suzek_uniref_2007} that  clusters homologous proteins  such that sequences within each cluster have at least 90\% identity to and 80\% overlap  with the longest sequence (seed). For inference of the Potts model, we used the default EV-Couplings settings for the rest of the parameters. Ultimately we were able to infer evolutionary models on 67 protein families corresponding to 67 chains in our testing set. These protein families amounted to 11,221 unique sites, for which we compare the EV-Couplings predictions on amino acid preferences with the H-CNN  predictions in Fig.~2F.

To measure the similarity between the predictions of H-CNN and EV Couplings, we  define the profile overlap $\rho$  for amino acid preferences at a given position. Specifically, for two sets of amino acid probabilities $P_{A},P_{B} \in [{\mathbb{R}^+_{\leq 1}}]^{19} $, the vector overlap is defined as the normalized dot-product between the centered probability vectors, 
\begin{equation}
	\rho = \frac{
		\big( P_A - \langle P_A \rangle_\text{sites}\big) \cdot \big( P_B - \langle P_B \rangle_\text{sites} \big)
	}{
		\sqrt{
			|P_A - \langle P_A \rangle_\text{sites} |^2 |P_B - \langle P_B \rangle_\text{sites} |^2
		}
	}
\end{equation}
where $\langle \cdot\rangle_\text{sites}$ denotes the average of  probability vector overs all  sites in the data.  Vector overlap takes values between -1 and 1, with 1 indicating a perfect alignment between the two (centered) vectors of amino acid preferences, and -1 indicating complete disagreement.

\section{Effect of shear perturbation on protein micro-environments}
We locally perturbed protein structures around their crystal conformation with the shearing deformation. As demonstrated in Fig.~3, shearing by an angle $\delta$ at a given residue of a protein is a perturbation to the backbone dihedral angles at two neighboring residues given by
\begin{align}
	(\phi_i,\psi_{i-1}) \longrightarrow (\phi_i + \delta,\psi_{i-1} - \delta)
\end{align}
This perturbation was chosen because it can substantially change the local protein structure  near the residue of interest while minimally affecting the far-away residues.

Physical interactions in a protein should be dependent on the pairwise distances between atoms. We define $D_{ab}$ to be the pairwise distance in angstroms between a pair of atoms $(a,b)$ in the protein structure. Then $\Delta D^{i,\delta}_{ab} = D^{i,\delta}_{ab} - D^0_{ab}$ measures the deviation of these pairwise distances from the   pairwise distances in the native (unperturbed) structure $D^0_{ab}$, when site $i$ is sheared by an angle $\delta$ (Fig.~3). We estimate the total shear-induced distortion in a protein structure  as the  root-mean-square of the deviations of pairwise distances $\text{RMS}\Delta D^{k,\delta}_{ij}$; here, the average is taken over all pairs of atoms that were at distances smaller than 10~\AA\, in the unperturbed structure, i.e.,  $\{(a,b) \, | \, D^0_{ab} < 10\, \text{\AA}\}$.
 
We characterize the model's response to the perturbation  by the change in the protein network energy (Fig.~3), defined as  the sum of pseudo-energies (logits) of amino acids over all sites in the protein,
\EQ
	E = \sum_{i \in \text{sites}} \tilde{E}_{i}^{\alpha_i}
	\label{eq.prot-en}
\EE
where $\alpha_i$ denotes the protein's amino acid at site $i$.

\begin{figure*}[p]
\includegraphics[width=\textwidth]{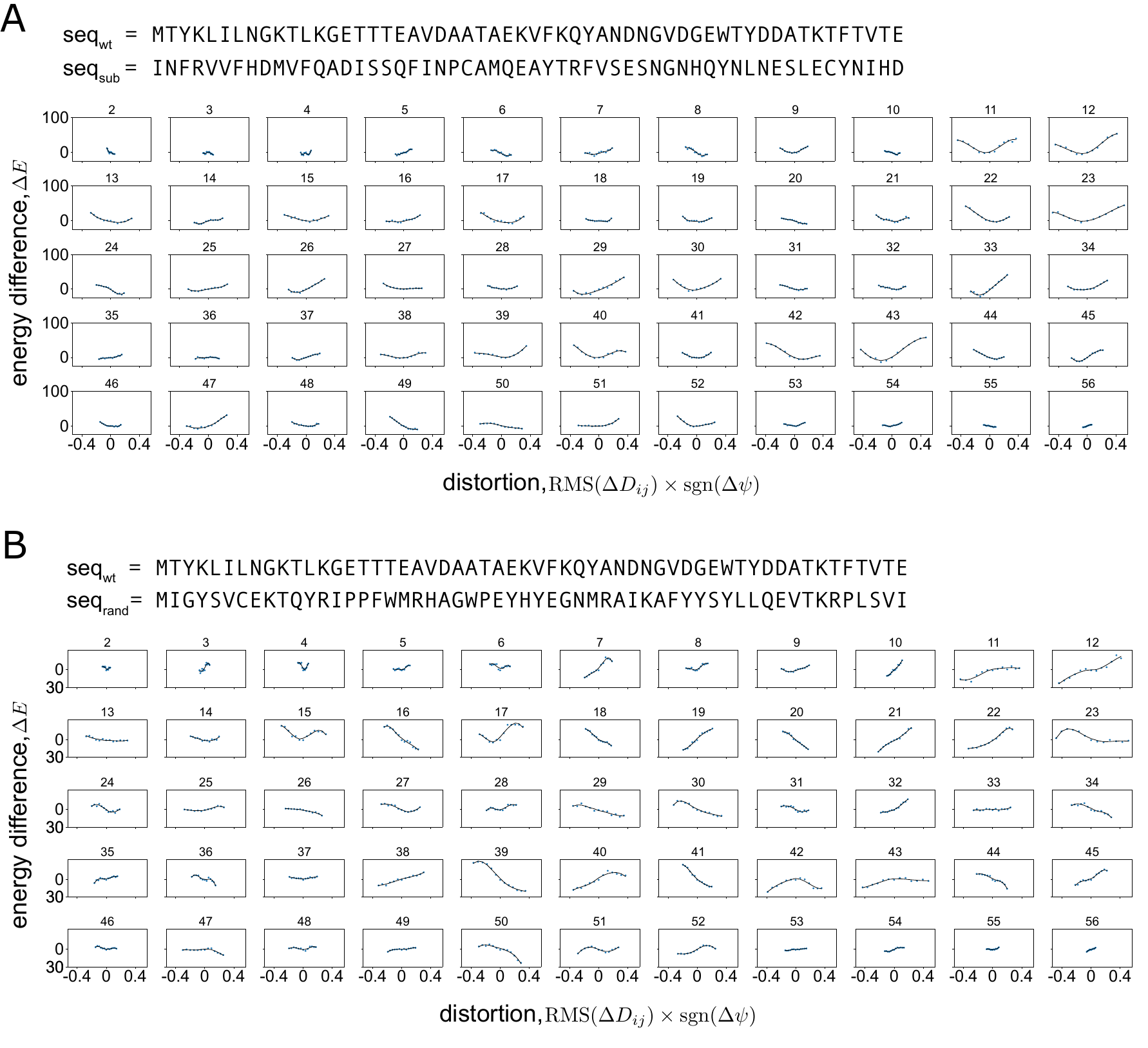}
\caption{
{\bf  Robustness of  response to shear perturbation.} 
Similar to Fig.~3, but the change in network energy is evaluated for alternative amino acids at the shear position, while keeping the  surrounding protein structure intact (i.e., using the wild-type neighborhood for model evaluation). The alternative amino acids are drawn according to the BLOSUM62 matrix  in {\bf (A)}, and randomly in {\bf (B)}; the wild-type and the alternative sequence are shown each panel. When amino acids are substituted according to their BLOSUM62 scores, potential wells are generally recovered (A), while no energy minima is recovered for random substitutions (B). Specific sequences used are given by $\text{seq}_{\rm sub}$ (substitutions according to the BLOSUM62 matrix), and $\text{seq}_{\rm rand}$ (random substitutions) in each panel and are shown in comparison to the wildtype sequence  $\text{seq}_{\rm wt}$.
\label{Fig:SI_shear}}
\end{figure*}

\begin{figure*}[p]
\includegraphics[width=\textwidth]{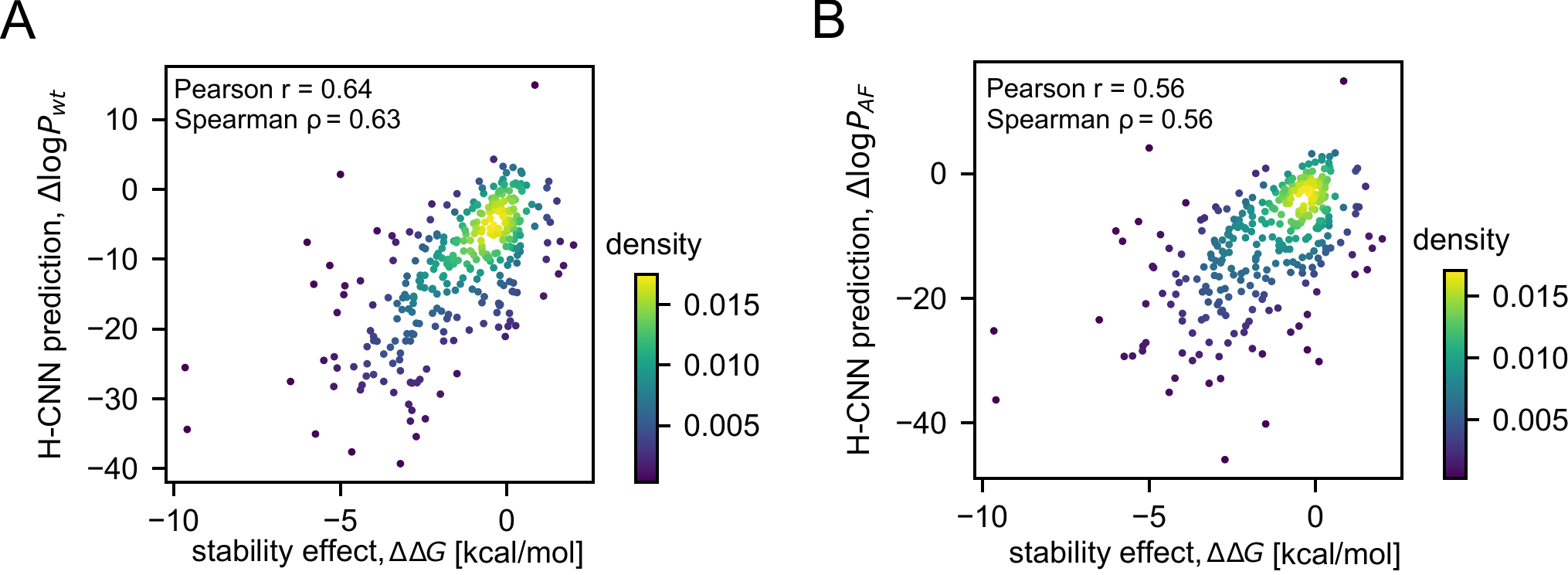}
\caption{
{\bf H-CNN predictions for the stability effect of all available single point mutations in T4 lysozyme.} 
 {\bf (A)} The true positive vs. false positive rates (ROC curve) is shown for classification of the 40 amino acid mutations (from Fig.~4A) into deleterious and neutral/beneficial classes based on the H-CNN predicted log-probability ratios  $\Delta \log P$, using the wild-type and the mutant structures  (orange), only the wild-type structure (purple),  the wild-type structure for the wild-type and an {\em in silico} relaxed structure for each mutant (green), and the AlphaFold predicted structure of the wild-type (blue). The corresponding area under the curves (AUCs)   are reported in the figure.
 {\bf (B)} H-CNN predictions for the relative log-probabilities using  the AlphaFold predicted structure $\Delta \log P_{\AF}$ are shown against the experimentally measured $\Delta \Delta G$ values for 310 single point mutation variants of  T4 lysozyme. Mean $\Delta\Delta G$ was used when multiple experiments reported values for the same variant. The colors  show the density of points in each panel as calculated via Gaussian kernel density estimation. H-CNN predictions correlate well with the experimental values, with the Pearson and Spearman correlations indicated in each panel.
\label{Fig:SI_T4_AF}
}
\end{figure*}

\begin{figure*}[p]
\includegraphics[width=\textwidth]{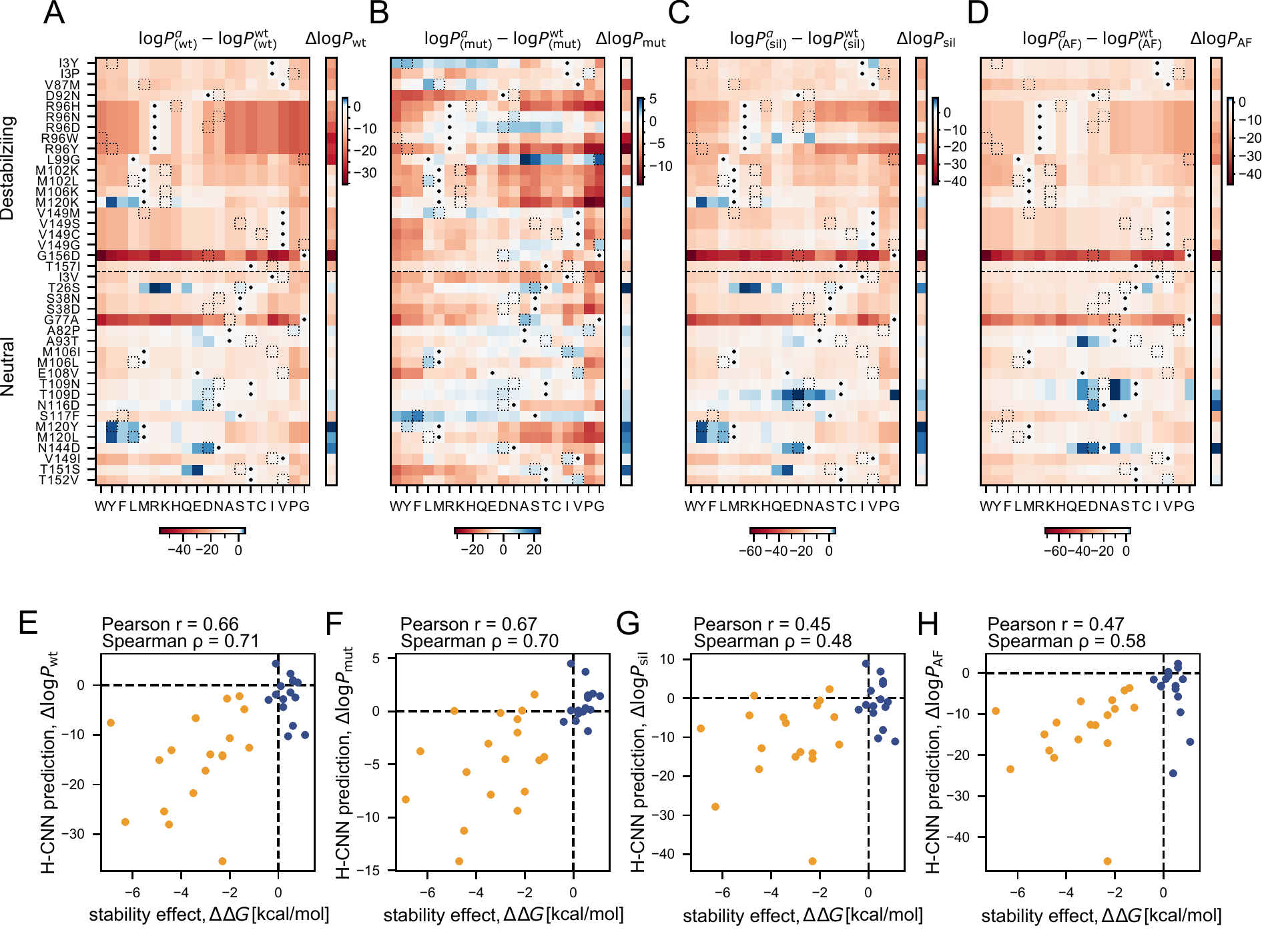}
\caption{
{\bf  Predictions for the stability effect of mutations in T4 lysozyme with different protein structures.} 
{\bf (A-D)} The large heatmaps show the H-CNN predicted log-probability of different amino acids relative to the wild-type at the substituted positions along the T4 lysozyme, indicated on the left-hand side of panel A (similar to Fig.~4A). In each heatmap, the dots indicate the identity of the wildtype amino acid, and the dotted squares indicate the amino acid in the specified position for the mutated variant in each row. The predictions for log-probabilities are evaluated using the wild-type structure~(A),  the mutant crystal structures~(B),  the wild-type structure relaxed {\em in-silico} via PyRosetta for the specified substitution~(C), and the computationally predicted structure of the wild-type  T4 lysozyme using AlphaFold~(D). The relative log-probabilities of the mutations of interest are shown separately in one column next to each heatmap. 
 They are calculated with the structural information that is available in different scenarios. For the wild-type, mutant, and {\em in silico} structures, $\Delta\log P_{*}=\log P^{\mut}_{(*)}/P^{\wt}_{(\wt)}$, with $*$ indicating the specified protein structure in each case. This definition is akin to $\Delta\Delta G$. For the AlphaFold structure, we use $\Delta\log P_{\AF}=\log P^{\mut}_{(\AF)}/P^{\wt}_{(\AF)}$---a quantity which could be evaluated in the absence of any experimental structural knowledge for a protein.  {\bf (E-H)} The predicted relative log-probabilities shown in panels (A-D) are plotted against the experimentally measured $\Delta\Delta G$ values for each variant. Destabilizing mutations are shown in orange and neutral/beneficial mutations are shown in blue.  Overall, H-CNN predictions correlate well with the mutational effects on protein stability, with the Pearson and Spearman correlations indicated in each panel and AUC values to discriminate between destabilizing and neutral mutations indicated in Fig.~4D. Accounting for changes in the local protein structures due to mutations in (B,~C)   sets the correct reference point in predictions such that the estimated log-probability ratio is overall positive for neutral/beneficial mutations and negative for destabilizing mutations.
\label{Fig:SI_T4}}
\end{figure*}

As shown in Fig.~3D, the native protein structure appears to be at the energy minimum of the network and shear perturbations often result in less favorable protein micro-environments. To assess the robustness of this result, we evaluated the protein network energy in eq.~\ref{eq.prot-en} for alternative  amino acids, while keeping the  surrounding protein structure intact (i.e., using the wild-type neighborhood for model evaluation). Fig.~\ref{Fig:SI_shear} shows that if amino acid substitutions preserve the physico-chemical properties of the residue (i.e., substitutions according to the BLOSUM62 matrix), the undistorted structure remains close to the  energy minimum, yet the landscape becomes shallower. However, for random amino acid substitutions, no energy minima is recovered. These results  further suggest that H-CNN has learned an effective physical potential for protein micro-environments.

%\nocite{*}
%\bibliographystyle{plos2015}
%\bibliography{bib_armita,SI_bib}

\newpage{}

\begin{table}[h!]
\begin{center}
\begin{tabular}{c | c | c | c | c | c | c}%\hline
Hyperparameter & variable & broad sampling space & sampling scheme & fine sampling space &  \thead{optimal fc \\network value }& \thead{optimal sc\\ network value}\\\hline%\hline
 batch size & $n_b$ & [8, 5000] & uniform & 256 & 256 & 256 \\%\hline
 dropout rate & $\lambda_{dr}$ & [0,0.6] & uniform &$ [0,10^{-5}]$ & $5.49 \times 10^{-4} $ & $1.95\times 10^{-2}$\\%\hline
 hidden dimension & $d_h$ & fc:[6,20], sc:[50,160] & uniform & fc:[8,20], sc:[100,500] & 14 & 109 \\%\hline
 No. CG layers & $n_l$ & [1,5] & uniform & [4,5]& 4 & 5\\%\hline
 No. dense layers & $n_{d}$ & [1,5] & uniform & [2,4] & 2 & 2\\%\hline
 Max spherical degree & $L_{max}$ & [2,7] & uniform & [4,5] & 5 & 5\\%\hline
 learning rate & $\lambda_l$ & $[10^{-6}, 10^{-1}]$ & exponential & $10^{-3}$ & $10^{-3}$ & $10^{-3}$\\%\hline
 regularization strength & $\lambda_r$ & $[10^{-6}, 10^{-1}]$ & exponential & $[10^{-10},10^{-16}]$ & $1.2 \times 10^{-16}$ & $1.89\times10^{-14}$ \\%\hline
\end{tabular} 
 \caption{\label{table:hyper}{\bf Hyperparameter bounds and optimal values.} Hyperparameters varied during model optimization are listed.  The two different bounds used during the two steps of hyperparameter optimization are shown in the broad and fine sampling spaces. Different bounds for fully connected (fc) and simply connected (sc) networks are shown when appropriate. Finally, the optimal value is listed for both the optimal fully connected and simply connected networks.\\\\
 }
\end{center}
\end{table}

\begin{table}[h!]
\begin{center}
\begin{tabular}{c | c | c | c | c | c | c | c}
	method & \thead{rotationally\\ invariant} & dataset & post-processing & N & No. of parameters & training time & accuracy \\ \hline
	H-CNN  & yes & ProteinNet & \thead{charge\\ hydrogen\\ SASA} & $2.8\times10^6$ & $3.6\times10^6$ & 4.54 hours & 68\% \\ \hline
	3D CNN \cite{Torng2017-aq} & no & SCOP \& ASTRAL & None & $7.2 \times 10^5$ & $10^7$ & 3 days & 40\% \\ \hline
	3D CNN \cite{Shroff2020-bu} & no & SCOP \& ASTRAL & \thead{PDB-REDO\cite{joosten_pdb_redo_2009}\\ charge\\ hydrogen\\ SASA} & $1.6 \times 10^6$ & $6.1 \times 10^7$ & -- & 70\% \\ \hline
	Spherical CNN \cite{boomsma_spherical_2017} & approx. & PISCES & \thead{charge\\ hydrogen} & -- & $6 \times 10^7$ & -- & 56\%\\ \hline
	Steerable CNN \cite{weiler_3d_2018} & yes & SCOP \& ASTRAL & \thead{PDB-REDO\\ charge\\ hydrogen\\ SASA} & $1.6 \times 10^6$ & $3.3 \times 10^7$ & -- & 58\%
\end{tabular}
\caption{\textbf{Comparison of structure-informed models for protein neighborhoods.} H-CNN and existing methods trained on classifying all-atom neighborhoods are listed along with the summary quantities of the models. H-CNN demonstrates the power of respecting symmetry since it has fewer parameters and trains faster than 3D-CNNs despite being trained on at least the same amount of data. \\\\
\label{tab:performance}}
\end{center}
\end{table}

\begin{table}[h!]
\resizebox{\textwidth}{!}{
\begin{tabular}{|c|c|c|c|c|c|c|c|c|c|c|c|c|c|c|c|c|c|c|c|c|}\hline
       variant & I3Y      & I3P      & V87M     & D92N     & R96H     & R96N     & R96D     & R96W     & R96Y     & L99G     & M102K    & M102L    & M106K    & M120K    & V149M    & V149S    & V149C    & V149G    & G156D    & T157I    \\ \hline
PDB id    & 1L18     & 1L97     & 1CU3     & 1L55     & 1L34     & 3CDT     & 3C8Q     & 3FI5     & 3C80     & 1QUD     & 1L54     & 1L77     & 231L     & 232L     & 1CV6     & 1G06     & 1G07     & 1G0P     & 1L16     & 1L10     \\ \hline
$\Delta\Delta G$~[kcal/mol]    & -2.3     &          & -2.3     & -1.4     &          & -3.0     & -3.5     & -4.5     & -4.7     & -6.3     & -6.9     & -2.11    & -3.4     & -1.6     & -2.8     & -4.4     & -2.0     & -4.9     & -2.3     & -1.2     \\ \hline
pH     & 6.5      &          & 5.4      & 5.7 -5.9 &          & 5.35     & 5.35     & 5.35     & 5.35     & 5.4      & 5.3      & 5.7      & 3        & 3        & 5.4      & 5.4      & 5.4      & 5.4      & 6.5      & 6        \\ \hline
source & ~\cite{Matsumura1988-gz} & ~\cite{Dixon1992-ye} & ~\cite{Gassner1999-uq}& ~\cite{Nicholson1991-wr} & ~\cite{Weaver1989-fo} & ~\cite{Mooers2009-ik} & ~\cite{Mooers2009-ik} & ~\cite{Mooers2009-ik} & ~\cite{Mooers2009-ik} & ~\cite{Wray1999-ox} & ~\cite{Dao-pin1991-gs} & ~\cite{Hurley1992-dp} & ~\cite{Lipscomb1998-yq} & ~\cite{Lipscomb1998-yq} & ~\cite{Gassner1999-uq}& ~\cite{Xu2001-vy} & ~\cite{Xu2001-vy} & ~\cite{Xu2001-vy} & ~\cite{Gray1987-hk} & ~\cite{Grutter1987-eo}\\\hline
\end{tabular}}
\phantom{0}
\resizebox{\textwidth}{!}{
\begin{tabular}{|c|c|c|c|c|c|c|c|c|c|c|c|c|c|c|c|c|c|c|c|c|}\hline
      variant & I3V      & T26S     & S38N     & S38D     & G77A     & A82P     & A93T     & M106I    & M106L    & E108V    & T109N    & T109D    & N116D    & S117F    & M120Y    & M120L    & N144D    & V149I    & T151S    & T152V    \\ \hline
PDB id    & 1L17     & 131L     & 1L61     & 1L19     & 1L23     & 1L24     & 129L     & 1P46     & 234L     & 1QUG     & 1L59     & 1L62     & 1L57     & 1TLA     & 1P6Y     & 233L     & 1L20     & 1G0Q     & 130L     & 1G0L     \\\hline
$\Delta\Delta G$~[kcal/mol]    & -0.4     &          & 0.6      &          & 0.4      & 0.8      &          & 0.2      & 0.5      & 0.7      & 0.1      & 0.6      & 0.6      & 1.1      & -0.1     & 0.5      &          & -0.1     &          & 0.2      \\\hline
pH   & 6.5      & 5.4      & 5.7 -5.9 & 5        & 6.5      & 6.5      & 5.4      & 5        & 3        & 5.4      & 5.7 -5.9 & 5.7 -5.9 & 5.7 -5.9 & 5.4      & 5        & 3        & 5        & 5.4      & 5.4      & 5.4      \\ \hline
source & ~\cite{Matsumura1988-gz} & ~\cite{Pjura1993-mx} & ~\cite{Nicholson1991-wr} & ~\cite{Nicholson1988-uu} &~\cite{matthews_enhanced_1987} &~\cite{matthews_enhanced_1987} & ~\cite{Pjura1993-mx} & ~\cite{Mooers2003-eq} & ~\cite{Lipscomb1998-yq} & ~\cite{Wray1999-ox} & ~\cite{Nicholson1991-wr} & ~\cite{Nicholson1991-wr} & ~\cite{Nicholson1991-wr} & ~\cite{Anderson1993-ap} & ~\cite{Mooers2003-eq} & ~\cite{Lipscomb1998-yq} & ~\cite{Nicholson1988-uu} & ~\cite{Xu2001-vy} & ~\cite{Pjura1993-mx} & ~\cite{Xu2001-vy}\\\hline
\end{tabular}}
\caption{\label{tab:T4Lys}{\bf T4 lysozyme variants.} Each of the 40 variants, shown in Fig.~4,  with the respective protein structure is listed along with the corresponding PDB entry. When available, the $\Delta\Delta G$ associated with a mutation and the pH of the experimental setup in which protein stability was measured are provided.}
\end{table}

\newpage{}

 \begin{figure*}[p]
\includegraphics[height=8.5in]{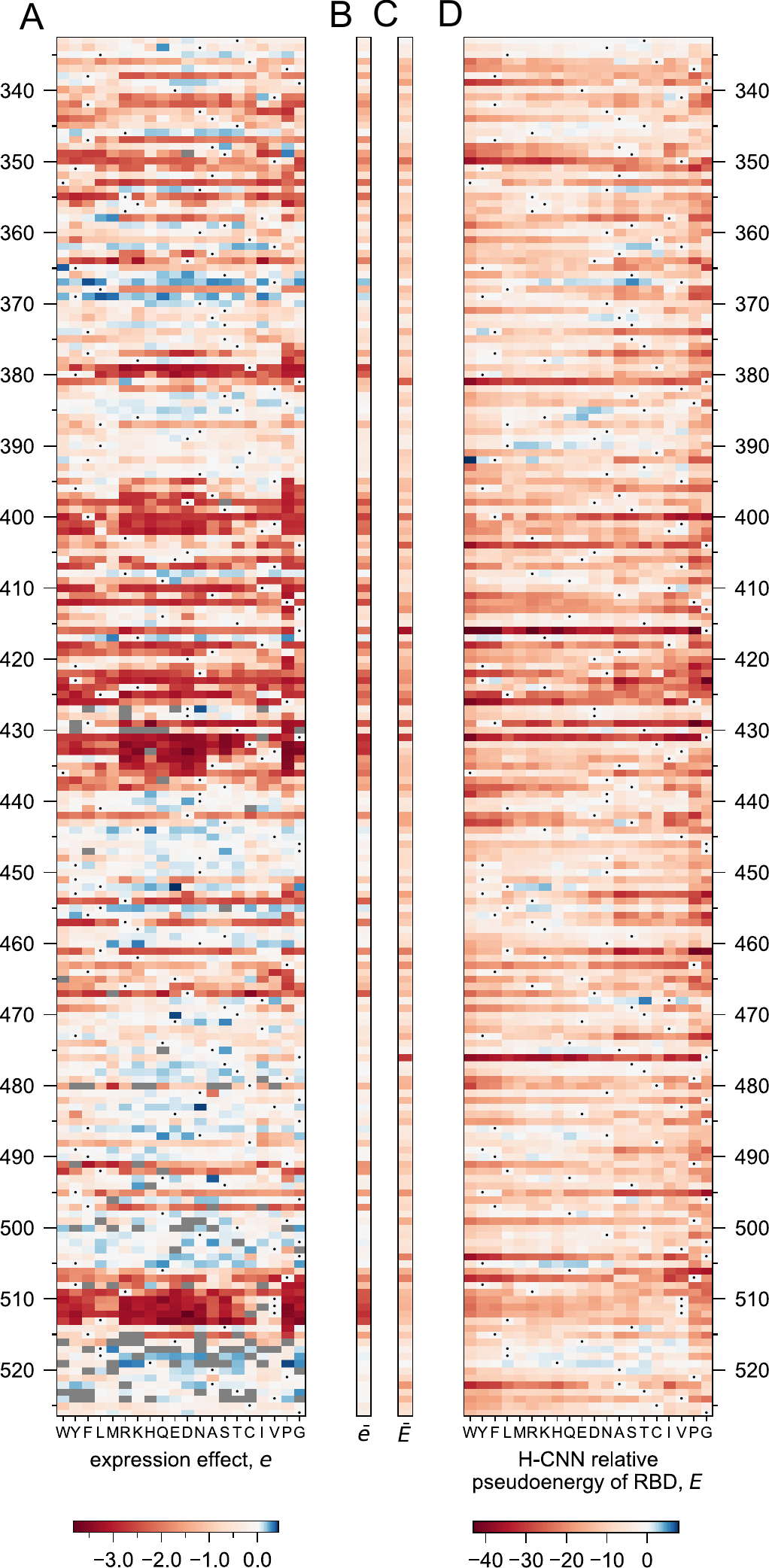}
\caption{
{\bf Experimental RBD expression of SARS-CoV-2 in DMS experiments and the H-CNN predictions.} {\bf (A)} Experimental expression of single-point mutation variants of  SARS-CoV2 RBD relative to the wildtype (dots) as measured using yeast surface display experiments. Expression effect $e$ is calculated as $\Delta\log_{10}(\text{MFI})$ where MFI is the mean fluorescent intensity. Mutations to G or P dominate the negative tail of expression effects. To better show the effect of all mutations, the negative portion of the colorbar is bounded by the expression of the most detrimental variant that is not a mutation to G or P. {\bf (B)} The mean expression effect of mutations $\bar{e}$ is listed next to each site. {\bf (C,D)} The  H-CNN  predicted stability effect of mutations (D), and the mean predicted stability effect for each site $\bar{E}$ (C), evaluated based on the  computationally isolated RBD structure, are shown. Again the negative portion of the colorbar is bounded from below by the most destabilizing prediction not involving mutations to G or P.  
\label{Fig:SI_expr}}
\end{figure*}

 \begin{figure*}[p]
\includegraphics[height=8.5in]{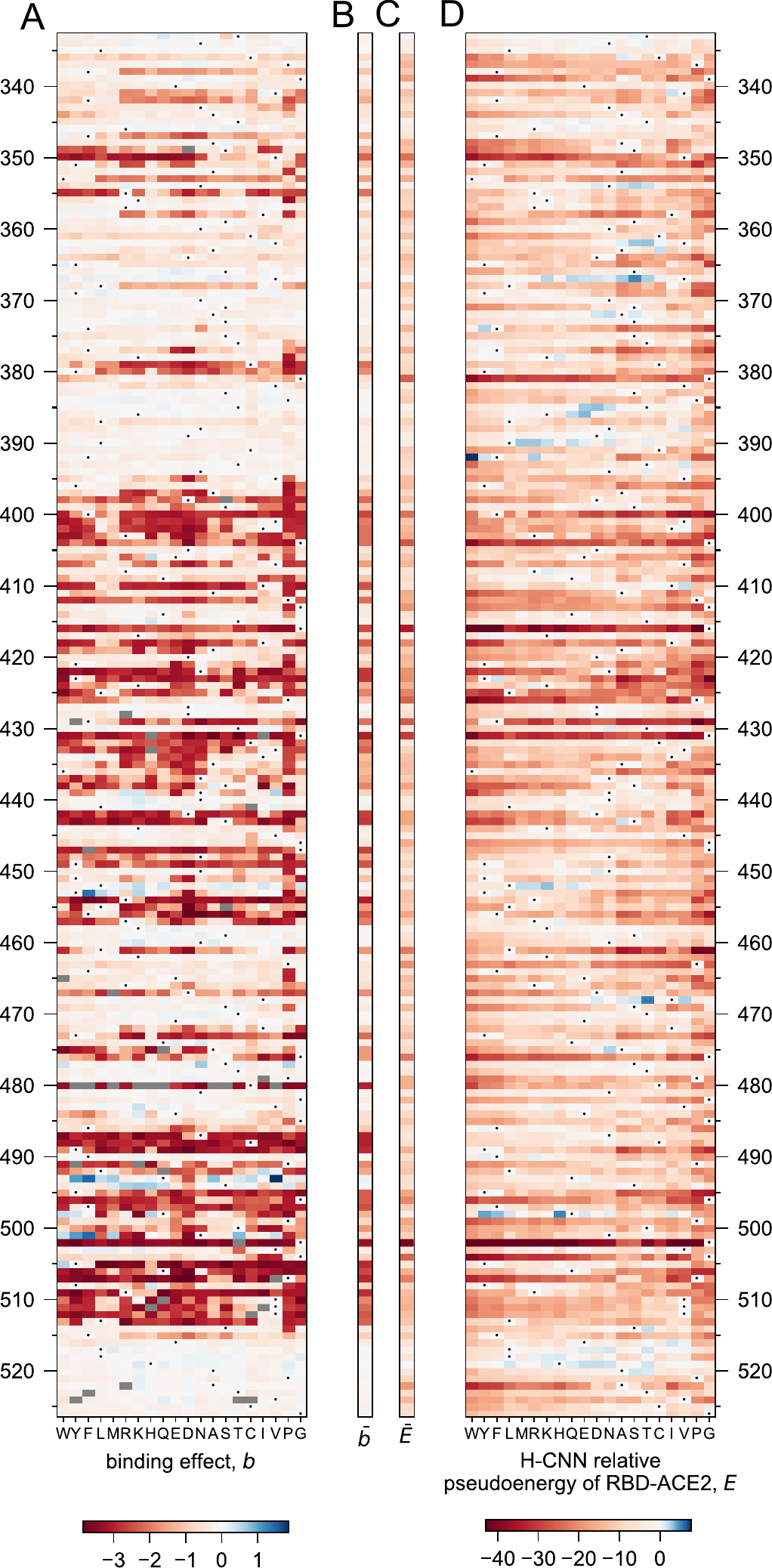}
\caption{
{\bf Experimental RBD-ACE2 binding affinity  in DMS experiments and the H-CNN predictions.} {\bf (A)} Experimental binding of single-point mutation variants of  SARS-CoV2 RBD to the human SCE2 receptor relative to the wild-type (dots) as measured using yeast surface display experiments. Binding effect $b$ is reported from $\Delta\log K_D$ where negative values represent weaker binding relative to the wildtype. Mutations to G or P dominate the negative tail of expression effects. To better show the effect of all mutations, the negative portion of the colorbar is bounded by the expression of the most detrimental variant that is not a mutation to G or P. {\bf (B)} The mean binding effect of mutations $\bar{b}$ is listed next to each site. {\bf (C,D)} The  H-CNN  predicted  effect of mutations (D) and the mean predicted stability effect for each site $\bar{E}$ (C), evaluated based on the  crystal structure of RBD-ACE2 complex, are shown. Again the negative portion of the colorbar is bounded from below by the most destabilizing prediction not involving mutations to G or P. 
\label{Fig:SI_bind}}
\end{figure*}

 \begin{figure*}[p]
\includegraphics[width=0.5\textwidth]{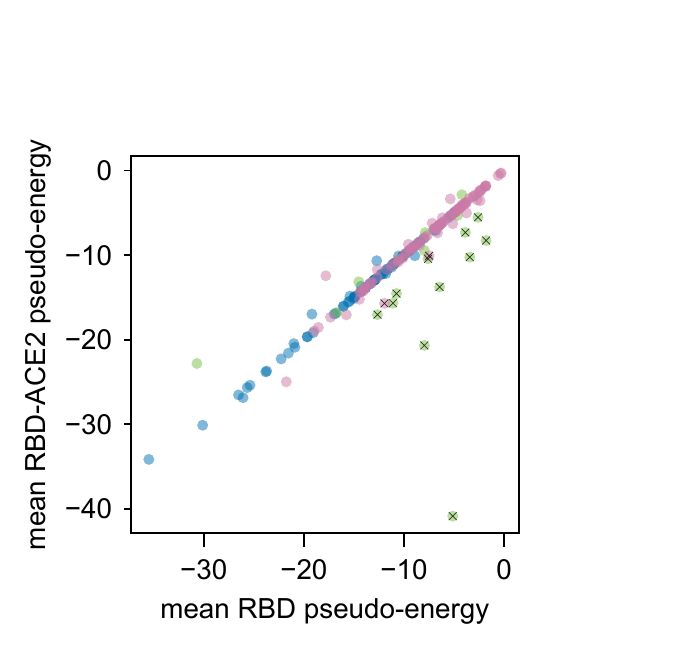}
\caption{
{\bf H-CNN predictions for stability and binding of RBD.} 
The H-CNN predicted mean  pseudo-energy per site predicted from the RBD-ACE2 protein complex (measure of binding) is plotted as a function of the mean  pseudo-energy per site, inferred from the isolated RBD structure (measure of stability). Points are colored according to their experimental values of expression and binding as illustrated in Fig.~5. Most points lie on the diagonal since the RBD pseudo-energy is determined from the RBD-ACE2 crystal structure with the ACE2 masked. However, sites at the RBD-ACE2 interface  show deviations from the diagonal. 
H-CNN prediction for sites that are  tolerant of mutations for stability but not binding are indicated by crosses (below the diagonal points with large stability values). These sites tend to be  at the RBD-ACE2 interface and overlap with the green category from Fig.~5.
\label{Fig:SI_DMS_site_pred}}
\end{figure*}